\documentclass[useAMS,usenatbib]{mn2e} 
\usepackage{epsfig}
\usepackage{xspace}
\usepackage{hhline}
\usepackage{amssymb}
\usepackage{amsmath}
\usepackage{multirow}
\usepackage{dcolumn}
\usepackage{verbatim}
\usepackage[figuresright]{rotating}
\usepackage{calc}
\usepackage{booktabs}
\usepackage{chngcntr}

\usepackage[%
  breaklinks=true,%
  colorlinks=true,%
  linkcolor=blue,anchorcolor=blue,%
  citecolor=blue,filecolor=blue,%
  menucolor=blue,pagecolor=blue,%
  urlcolor=blue]{hyperref}

\newcolumntype{.}{D{.}{.}{-1}}

\newcommand{\chandra}{{\it Chandra}\xspace}

\newcommand{\HST}{{\it HST}\xspace}

\newcommand{\wavdetect}{{\it wavdetect}\xspace}
\newcommand{\wvdecomp}{{\it wvdecomp}\xspace}

\newcommand{\sS}[1]{\mbox{$\rm{}^{#1}$}}
\newcommand{\Ss}[1]{\mbox{$\rm{}_{#1}$}}

\newcommand{\nHt}{\mbox{$N_{\mbox{\scriptsize H22}}$}\xspace}

\newcommand{\Deg}{\mbox{$^\circ$}\xspace}
\newcommand{\x}{\mbox{$\times$}}

\newcommand{\Ha}{\mbox{$H_\alpha$}\xspace}

\newcommand{\lcgs}{\mbox{erg s\sS{-1}}\xspace}
\newcommand{\fcgs}{\mbox{erg s\sS{-1} cm\sS{-2}}\xspace}
\newcommand{\sbcgs}{\mbox{erg s\sS{-1} cm\sS{-2} deg\sS{-2}}\xspace}

\newcommand{\F}{{\it F}\xspace}
\newcommand{\VF}{{\it VF}\xspace}

\renewcommand{\topmargin}{-50pt} 

\renewcommand{\textfloatsep}{10pt}
\renewcommand{\dbltextfloatsep}{10pt}

\title[Dominant MCVs in GRXE]{Dominance of Magnetic Cataclysmic Variables 
in the Resolved Galactic Ridge X-ray Emission of the Limiting Window}

\author[J. Hong]{
JaeSub Hong\sS{1}\thanks{E-mail: jaesub@head.cfa.harvard.edu} 
\\
\sS{1}Harvard-Smithsonian Center for Astrophysics, 
60 Garden St., Cambridge, MA 02138 }

\begin{document}

\maketitle
\counterwithout{figure}{section}

\begin{abstract} 

The diffuse appearance of the Galactic Ridge X-ray Emission (GRXE) 
has been puzzling since its discovery due to lack of compelling theories for 
sustainable hot
diffuse X-ray emission in the Galactic plane.
Recently \citet{Revnivtsev09} 
(R09) claimed that $\sim$90\% of the 6.5--7.1 keV X-ray flux
from a small section of a low
extinction region at 1.4\Deg\ south of the Galactic
Center has been resolved to discrete sources with $L\Ss{X,\  2-10\  keV}
\gtrsim 4\x 10^{-16}$ \fcgs, using ultra-deep (1 Ms) \chandra ACIS-I
observations.  They also concluded that coronally active stars such
as active binaries (ABs) contribute $\sim$60\% of the resolved flux.
However, our recent discovery of a large population of magnetic
cataclysmic variables (MCVs) in the same region suggests their significant
role in the resolved hard X-ray flux.  In addition, deep X-ray surveys of
other several Galactic Bulge fields over the past decade have indicated
that MCVs are likely the major contributor in the hard X-ray emission above
2--3 keV.  To solve this mystery, we have conducted an independent
indepth analysis of discrete X-ray sources in the low extinction region.
The total fraction of the 6.5--7.1 keV flux we can confidently claim as
resolved is $\sim$70--80\%, which largely agrees with R09 but leaves some
room for diffuse components.  However, despite the various attempts,
we consistently find that the resolved hard X-ray flux above 3 keV
is dominated by relatively bright, hard X-ray sources such as MCVs,
whereas the contribution from relatively faint, soft sources such as ABs
is below 20\%.  We describe in detail our analysis procedure in order
to elucidate possible origins of the discrepancy.

\end{abstract}

\section{Introduction}

The X-ray glows along the Galactic plane, discovered almost 30 years ago,
form a narrow continuous X-ray bright ridge, known as the Galactic Ridge
X-ray Emission (GRXE)  
\citep{Worrall82, Warwick85, Koyama86}.  
The origin of the GRXE, whether truly diffuse or
from unresolved discrete sources, has been debated ever since.  
The GRXE resembles
X-ray emission from an optically thin plasma of a high temperature (a
few keV) with emission lines from highly ionized heavy elements such
as Si and Fe \citep{Koyama86,Yamauchi93}.  However, the shallow
Galactic gravity and lack of energy
source to sustain such a plasma suggest discrete sources as the origin
of the GRXE \citep{Worrall83, Yamauchi96, Kaneda97}.

\begin{table*}
\begin{minipage}{0.99\textwidth}
\caption{\chandra ACIS-I observations of the LW}
\begin{tabular*}{0.99\textwidth}{c@{\extracolsep{\fill}}ccccrrcccl}
\hline
Obs.	& Start Time		&R.A.	& Decl.		& Offset\sS{a}	& Roll		& Exposure	&GTI\sS{b}	& Obs.		& Chips\\
ID	& (UT: y/m/d h:m)	&(\Deg)	& (\Deg)	& (\arcmin)	& (\Deg)	& (ks)		&(ks)		& Mode 		&	\\
\hline
6362	&2005/08/19 16:15	&267.86694  &-29.59235	& 0.1		& 273		& 38 		&37.7 		& {\ \F}		&0123..67..\\
5934	&2005/08/22 08:16	&267.86692  &-29.59233	& 0.1		& 273		& 41 		&{30.8} 	& {\ \F}		&0123..67..\\
6365	&2005/10/25 14:55	&267.86630  &-29.59212	& 0.1		& 265		& 21 		&20.7 		& {\ \F}		&0123..67..\\
9505	&2008/05/07 15:29	&267.85740  &-29.57123	& 1.3		& {82}		& 11 		&10.7		& \VF		&0123..6...\\
9855	&2008/05/08 05:00	&267.85741  &-29.57124	& 1.3		& {82}		& 57 		&55.9		& \VF		&0123..6...\\
9502	&2008/07/17 15:45	&267.86685  &-29.59108	& -		& 281		& 167 		&164.1		& \VF		&0123..6...\\
9500	&2008/07/20 08:11	&267.86148  &-29.58793	& 0.3		& 280		& 165 		&162.6		& \VF		&0123..6...\\
9501	&2008/07/23 08:13	&267.86399  &-29.58953	& 0.2		& 279		& 135 		&131.0		& \VF		&0123..6...\\
9854	&2008/07/27 05:53	&267.87404  &-29.59630	& 0.5		& 278		& 25 		&22.8 		& \VF		&0123..6...\\
9503	&2008/07/28 17:37	&267.86852  &-29.59311	& 0.2		& 275		& 103 		&102.3		& \VF		&0123..6...\\
9892	&2008/07/31 08:07	&267.86853  &-29.59312	& 0.2		& 275		& 65 		&65.8 		& \VF		&0123..6...\\
9893	&2008/08/01 02:44	&267.87098  &-29.59490	& 0.3		& 275		& 45 		&42.2 		& \VF		&0123..6...\\
9504	&2008/08/02 21:23	&267.87097  &-29.59490	& 0.3		& 275		& 127 		&125.4		& \VF		&0123..6...\\
\hline
\end{tabular*}
Notes. 
\sS{a}The aim point offset relative to Obs.~ID 9502.
Table 1 in H12 shows the target coordinates of each pointing.
\sS{b}The Good Time Intervals (GTIs) are selected by
the fluctuations ($<3\sigma$)
of the background event rates in the 2.5--7 keV band, which is calculated
after the discrete source contribution is removed.
The GTIs shown here are based on the events processed through the EDSER
routine and the \VF mode cleaning (see \S\ref{s:vfmode}).  
An additional manual inspection ensures the removal of the
background flaring periods.  As a result, the selected GTIs are
slightly different from those in Table 1 in H12 (see the text).
\end{minipage}
\label{t:obs}
\end{table*}

With subarcsec angular resolution and superb sensitivity, the \chandra
X-ray Observatory launched a decade ago brought a new hope of revealing
the nature of the GRXE.
The early studies by \citet{Ebisawa01, Ebisawa05} using deep \chandra
observations (100 ks each) of two adjacent Galactic plane fields at
($l$, $b$) = (28.5\Deg, -0.2\Deg) showed
the GRXE in the 2--10 keV band is largely unresolved
with discrete sources of $L\Ss{X,\ 2-10\ keV} \gtrsim 3\x10^{-15}$ \fcgs.
They claimed the unresolved X-ray flux cannot be explained by the
known types of undetected, fainter discrete X-ray sources, suggesting a
significant portion of the GRXE is truly diffuse.
However, \citet{Revnivtsev07} claimed the GRXE in the same region
can be explained by discrete sources when taking into account 
a large number of unresolved, faint, coronally active X-ray sources
such as active binaries (ABs).
Motivated by the resemblance of the
Galactic distribution of the GRXE and the near
infrared (nIR) emission measured by Spitzer that closely follows stellar
population \citep{Revnivtsev06}, 
\citet[hereafter R09]{Revnivtsev09} conducted ultra-deep observations (1 Ms) of
a low extinction region (Limiting Window; LW)\footnote{R09 call this
region the 1.5\Deg Window.} at 1.4\Deg\ south of the
Galactic Center.
They claimed that 88 $\pm$ 12\% of the highly ionized iron emission line
(6.5--7.1 keV) in the GRXE is resolved to discrete sources and the known
contribution of the Cosmic X-ray Background (CXB).  Again they claimed
a large fraction ($\sim$60\%) of the resolved flux is from relatively
faint, soft coronal X-ray sources such as ABs.

Meanwhile, several surveys have been conducted to study faint X-ray
sources in the Galactic plane and Bulge fields in order to understand their 
composition, which can lead to clues of formation and
evolutionary history of X-ray sources as well as Galactic Bulge and Galaxy
\citep[e.g.][hereafter M03, M09, H09b]{Muno03,Muno09,Hong09b}. 
The studies of newly discovered several
thousands of Bulge X-ray sources in these fields 
have indicated that the hard X-ray emission is dominated by
low luminosity ($L_X \lesssim 10^{32-33} $ \lcgs) hard X-ray sources and that
their major candidates are likely
magnetic cataclysmic variables (MCVs)
\citep[M09, H09b]{Laycock05}. 

The LW is one of the low extinction Bulge fields, and the first deep
\chandra (100 ks) exposure of the region was conducted as a part of our
survey to understand the population of
Galactic Bulge X-ray sources \citep[H09b,][hereafter V09]{Berg06,Berg09}.
Our recent discovery of periodic sources in the LW also indicates
a large population of MCVs in the region and 
their significant role in the resolved hard X-ray flux
\citep[hereafter H12]{Hong12a}. A recent spectral analysis of the GRXE
using the {\it Suzaku} observations also shows the dominance of the MCVs
in the hard X-ray band of the GRXE \citep[see also Morihana et
al.~2012]{Yuasa12}.

Motivated by this new puzzle in source composition of the resolved
GRXE and their contribution in the hard X-ray flux, we have conducted
an independent indepth analysis of X-ray sources found in the combined
ultra-deep \chandra exposure of the LW.  In particular, the advanced
source search routine employed by R09 seems to nearly double the source number
compared to other conventional methods (\S\S\ref{s:search} \&
\ref{s:detection}).  Despite the differences among the source search
routines, the lack of studies comparing the results in the past also
motivates us to conduct this comparative study of the source search
routines for future analysis of other fields.
We summarize the observations (\S\ref{s:obs}) and 
describe in detail our
analysis procedure (\S\ref{s:analysis}).  We present
our results in comparison with R09 (\S\ref{s:results}): the possible
origins of the discrepancy are presented in the Appendix.  We also discuss
the implications of our findings for future studies
(\S\ref{s:discussion}).

\section{Observations} \label{s:obs}

The LW was observed for a total of 1 Ms exposure (100 ks in 2005 and 900
ks in 2008) with the \chandra ACIS-I instrument. Table~\ref{t:obs} shows
the basic observational parameters (see also R09 and Table 1 in H12).
The observations in 2005 were conducted in {\it Faint} mode (\F mode) with 6
ACIS CCDs enabled, and the rest in {\it Very Faint} mode (\VF mode) with 5 ACIS
CCDs enabled.  The \VF mode allows an additional background reduction
(see \S\ref{s:vfmode}) \citep{Vikhlinin02}. 
The CCD readout time and thus the correction factor for the readout background
depend on the number of enabled CCDs (e.g.~41 ms readout time with
3.2s frame time for 6 CCDs) \citep{Markevitch00}.
The aim points of the observations varied about
0.1\arcmin\ to 1.3\arcmin\ from each other.
The roll angles of two observations
($\sim$70 ks) in 2008 were about 200\Deg off from the rest.  The roll
angle and aim point variations made the CCD gaps less prominent in the
merged data set.

\begin{table*}
\centering
\begin{minipage}{0.99\textwidth}
\caption{Key Parameters of Various Analysis Approaches (the Default Options in Bold) }
\begin{tabular*}{0.99\textwidth}{l@{\extracolsep{\fill}}lllll}
\hline
Analysis Stages						& Parameters 		& Options Tested \& Results Quoted in This Paper (Bold)	\\
\hline
Event Process						& Pixel Repositioning;  &{\bf 1)~EDSER; Yes}	\\
(\S\ref{s:vfmode})					& \VF Mode Cleaning	& 2)~EDSER; No 	\\
							&			& 3)~Random; No*\\
							&			& 4)~None; Yes	\\
\hline
Event Merge 						& Boresight Correction; & {\bf 1)~Yes; Obs.~ID 9502} \\
(\S\ref{s:merge})					& Reference Projection	& 2)~Yes; Exposure Weighted Average \\
							&			& 3)~No; Obs.~ID 9502 \\
\hline
Source Search 						& Search Method; 	&{\bf 1)~wavdetect; 10\sS{-6}} \\
(\S\S\ref{s:search} \& \ref{s:detection})		& Threshold		&{\bf 2)~wvdecomp; 4.5$\sigma$} \\
							& 			&{\bf 3)~wvdecomp; 4.0$\sigma$} \\
							&			&{\bf 4)~R09; 4.0$\sigma$}\\
\hline
Aperture Photometry					& Aperture Size; 	& {\bf 1)~Fixed 2\arcsec radius; the Rest of HRES }\\
(\S\S\ref{s:ap}, \ref{s:ap_results} \& \ref{s:ap_res2})	& Background Region 	& 2)~95\% PSF at 1.5 keV; an Annulus around Each Source  	\\
\hline
Instrumental Background Model				& Data Set for  	& {\bf 1)~Stowed data (Period E)}, exposure: 367 ks \\
(\S\S\ref{s:instbkg} \& \ref{s:resolved})		& Modeling 		& 2)~Stowed data (Period D$+$E), exposure: 235$+$367 ks \\
\hline
\end{tabular*} \\
*Standard CXC Level 2 event file 
\label{t:analysis}
\end{minipage}
\end{table*}

\section{Data Analysis} \label{s:analysis}

The analysis procedures for discrete sources can be grouped into four
main steps: event processing \& selection, stacking, source search, and
aperture photometry. Calculation of the total resolved X-ray flux requires
modeling of the instrumental background in the analysis region.
We have created a new analysis pipeline based on a custom analysis
pipeline developed over the years for the \chandra Multi-wavelength Plane
(ChaMPlane) survey \citep{Grindlay05}\footnote{For some of the more recent
survey results, see also \citet{Servillat12,Berg12a}.}. The latter is
described in
detail in \citet[H09b]{Hong05}\.  The new analysis pipeline is built on version
4.3 CIAO analysis tools, and has many improvements over the previous
pipeline.  In particular, we have implemented various analysis 
approaches with multiple parameter choices including the one similar
to what was employed by R09 for comparison.
Table~\ref{t:analysis} summarizes key parameters of analysis approaches
used in this paper.

\subsection{Event Process} \label{s:vfmode}

The standard CXC\footnote{http://cxc.harvard.edu} pipeline provides
Level 1 and 2 event files of each observation. Over the years, there
are many subtle or significant improvements proposed and implemented
for event processing.  In order to identify the effects of these new
implementations, we reprocessed the Level 1 event files with a few
different options. Two main options considered for reprocessing are
pixel repositioning and \VF mode cleaning. The pixel repositioning is
introduced to reduce pixellation-induced artifacts.  For instance, the
Energy-Dependent Subpixel Event Repositioning (EDSER) routine is shown
to noticeably improve the Point Spread Function
(PSF) \citep{Li04}. The \VF mode cleaning reduces instrumental background events
by utilizing the 5\x5 pixel readout of the \VF mode \citep{Vikhlinin02},
but it may also remove some valid events.
The Level 2 event files from the standard CXC procedure are 
generated with the random pixel repositioning under no \VF mode cleaning,
which is equivalent to the \F mode data.  We have implemented
four approaches as shown in Table~\ref{t:analysis}.  Our default choice uses
the EDSER routine and the \VF mode cleaning for the 2008 observations.

Another minor,
but noticeable improvement from the earlier analysis (Table 1
in H12) is in the Good Time Interval (GTI) selection shown in
Table~\ref{t:obs}.
The GTIs are calculated to screen out events
acquired during background flares. We consider an interval good if
the background rate of the interval is $<3\sigma$ from the mean rate.
In this analysis, the background rates were calculated from events in 
the 2.5--7
keV band in 1 ks bins, which is known to be optimal for identifying
background flares \citep{Markevitch03}\footnote{\citet{Markevitch03}
recommend using the events of S3 chip in the 2.5--7 keV band for
flare identification, but S3 chip was not on for some observations, so
we use the events of each CCD in the same energy band.}.
In addition, they were generated after the
point source contribution was removed, based on a preliminary source
detection by the \wavdetect algorithm \citep{Freeman02}.  The point
source removal keeps the GTI selection algorithm from misrecognizing
bright flares of discrete sources as instrumental background flares.
For instance, we recovered an erroneously removed 20 min GTIs in Obs.~ID
9502 \citep{Hong12b}.  Finally, we also manually verified all the GTIs for any
anomaly. As a result, we have removed the full portion ($\sim$ 10 ks)
of a background flare in Obs.~ID 5934 from the GTIs, only a part of
which was identified by the automatic
procedure.  

\subsection{Event Merge} \label{s:merge}

We merge the selected events of multiple observations for the full
benefit of the ultra-deep exposure.  We mainly consider two options in
merging: boresight offset correction and the choice of
reprojection
tangential point.  Since the aspect and pointing errors of the \chandra observations can
be as large as 0.6\arcsec\ (90\% confidence)\footnote{\url{http://cxc.cfa.harvard.edu/cal/ASPECT/celmon}}, we use relatively bright X-ray sources detected
in each observation for boresight correction \citep{Zhao05,Hong05}. 
Relative to 
Obs.~ID 5934, the calculated boresight offsets of other
observations range from 0.1\arcsec\ to 0.34\arcsec. We merged the data
with and without the boresight offset correction for comparison.

For the default choice of the common reprojection tangential point, we use
the aim point of Obs.~ID 9502 with the longest exposure.  Since the
aim points varied as much as 1.3\arcmin\ from pointing to pointing
(Table~\ref{t:obs}),
we also used the exposure-weighted average aim point of the observations
for the reprojection tangential point for comparison.

\subsection{Source Search} \label{s:search}

The \wavdetect routine based on the wavelet algorithm \citep{Freeman02}
is one of the popular source detection tools in X-ray astronomy.
Its performance has been extensively studied and tested
\citep[e.g. M03,][]{Kim07}.
We have used the CIAO \wavdetect routine for source detection for many
applications in the past \citep[e.g.][H09b]{Hong05}.  The \wavdetect
routine delivers a list of sources with detection significance at
a somewhat conservative level under the standard parameter setting.
Many researchers have developed new techniques or improved the algorithm
to catch relatively faint sources missed by the standard \wavdetect
routine, some of which are visually identifiable even in the input images. 
These include the \wvdecomp routine by
A.~Vikhlinin\footnote{\url{http://hea-www.harvard.edu/RD/zhtools}}, 
a multi-statistics based approach 
by \citet{Wang04} and an enhanced \wavdetect algorithm  for multiple
observations by \citet{Kashyap11}.  In particular, the \wvdecomp algorithm is well suited
for detecting faint sources by implementing successive iterations that
remove the contribution from bright sources in the image.  A typical
improvement acquired by these new tools is a $\sim$ 10--30\% increase in
source number under the recommended parameter settings.  For instance,
M09 found additional $\sim$ 26\% of sources from the \wvdecomp routine
compared to the sources discovered by the \wavdetect routine (see \S\ref{s:detection}).

Since the final source number counting depends sensitively on some of
the input parameters of each detection method, we employ three source
detection lists along with the source list from R09 for subsequent
aperture photometry.
For the \wavdetect routine, we employ a typical threshold of $10^{-6}$,
which allows about one false source in each ACIS chip (1024 \x 1024
pixels, see also \S\ref{s:detanal} \& Appendix~\ref{s:detsig}).  For the \wvdecomp routine, we have tried two settings of the
significance threshold (4.5$\sigma$ or 4.0$\sigma$), which is the main
driver of the final source number count.  The original prescription
of the routine  recommends the threshold setting at 4.5$\sigma$, but
R09 lowered it to 4.0$\sigma$ for their analysis under the
following two justifications. First, sources only in the high resolution
(HRES) region (a central circular region of 2.56\arcmin\ radius) with
the highest sensitivity and finest spatial resolution, were considered
for analysis.  Second, a similar detection run on the instrumental
background data produces only 1 or 2 false sources.  We will discuss
the latter again in
\S\ref{s:detection}.  The number of iterations were fixed at 5 since there
is practically no change in source number after the fifth iteration.

\begin{table*}
\begin{minipage}{0.99\textwidth}
\caption{Source Number Counting in the HRES region (2.56\arcmin\
radius) of the LW}
\begin{tabular*}{\textwidth}{lcccccccccc}
\hline
(1)                                         & (2)        & (3)                                  & (4)        & (5)                                                   & (6)            & (7)                  & (8)        & (9)                               & (10)       \\
Search                                      & Source     & \multicolumn{2}{c}{Sources with}              &\multicolumn{4}{c}{Comparison: Sources not found in}                                             & \multicolumn{2}{c}{Source Number}           \\
\cmidrule(r){5-8} \cmidrule(l){9-10}Routine & Number     & \multicolumn{2}{c}{false det. prob.}          &\wvdecomp                                            & \wvdecomp      & \wvdecomp            & Ref Coord. &                                   & Stowed E   \\
\cmidrule(l){3-4}                           & 0.5--7 keV & $<$1\%                               & $<$0.005\% & $\ge4.5\sigma$                                        & $\ge4.0\sigma$ & $\ge4.0\sigma$ (R09) & Change     & 9--12 keV                         & 0.5--7 keV \\
\hline
\wavdetect                                  & 274        & 274                                  & 266        & 2 (1\%)                                               & 0 (0\%)        & 5 (2\%)              & 13 (5\%)   & 0                                 & 0          \\
\hline\wvdecomp                             &            &                                      &            &                                                       &                &                      &            &                                   &            \\
\ \ $\ge4.5\sigma$                          & 356        & 355                                  & 316        & --                                                    & 0 (0\%)        & 26 (7\%)             & 29 (8\%)   & 2                                 & 3          \\
\ \ $\ge4.0\sigma$                          & 439        & 429                                  & 346        & 83 (19\%)                                             & --             & 75 (17\%)            & 41 (9\%)   & 24                                & 25         \\
\ \ $\ge4.0\sigma$ (R09)                    & 473        & 456                                  & 337        & 143 (30\%)                                            & 109 (23\%)     & --                   & N/A        & N/A                               & N/A        \\
\hline
\end{tabular*}

\label{t:detection}
Notes. (2) The number of sources detected. 
(3) The number of sources with false detection probability 
$P_r$$<$1\% (or detection confidence $C$$>$99\%). (4) 
$P_r$$<$0.005\% (or $C$$>$99.995\%)
(see \S\ref{s:detanal}).
(5), (6) \& (7) The number of
unique sources not found by the other search methods in comparison.
(8) The number of unique sources compared to the case where the
reprojection tangential point was set at the exposure-averaged
aim point of the 13 observations instead of the aim point of Obs.~ID 9502.
(9) The number of sources detected from the image of the LW in the
9--12 keV band.
(10) The number of sources from the image of the reprojected Stowed
data set (Period E) in the 0.5--7 keV band.
\end{minipage}
\end{table*}

\subsection{Aperture Photometry of Discrete Sources} \label{s:ap}

Aperture photometry can also be applied in many different ways.  
For the aperture of a
given source, we often use a circle around the source position enclosing 95\% of PSF for 1.5 keV
X-rays, and for the background region,
an annulus with the inner and outer radii of 2\x\ and 5\x PSF respectively. The
background annulus region excludes the source regions of neighbors.
This aperture choice allows aperture photometry over the entire field,
even outside HRES.  R09 used a fixed 2\arcsec\ radius circle around each source
for the source aperture and the rest of the HRES region (excluding all the
source regions) for the background region.  The large, fixed background
region provides higher statistics for background counts, but it may not
properly reflect a local variation of the background around each source.
We employ both approaches for comparison (Table~\ref{t:analysis}).

For background subtraction, we need to know the aperture ratio of the
source to background regions. There are a few
ways to calculate this ratio; one is a simple geometric ratio of the
regions, which is more appropriate for dealing with internal instrumental
background, and another is an exposure-map (effective area) corrected
geometric ratio of the regions, which is more appropriate for diffuse
sky X-ray background.  Since both ratios are consistent within less
than a percent of each other,  here we use the ratio of the exposure-map
corrected areas, where the exposure map was generated for 1.5 keV X-rays.
Note although the effective area depends sensitively on energies, the
ratio of the source to background regions for aperture photometry
hardly does. We will discuss the ratio in more detail in
\S\ref{s:resolved} \& Appendix~\ref{s:ratio}.

\begin{figure} \begin{center}
\includegraphics*[width=0.47\textwidth,clip=true,trim=5mm 2mm 5mm 2mm]{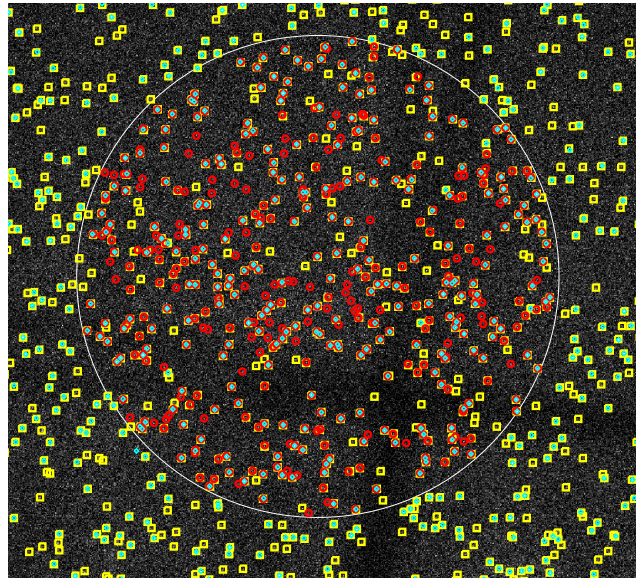}
\caption{The image of the HRES region of the LW marked with sources.
The (cyan) diamonds are from the \wavdetect routine, the (yellow)
squares from the \wvdecomp (4.0$\sigma$), and the (red) circles from R09. See Table~\ref{t:detection}.}
\label{f:image}
\end{center}
\end{figure}

The frequency of overlapping aperture regions increases,
as the source number count increases. 
In order to
avoid double counting of events due to the overlap we uniquely assign each
event in the overlapping regions to a source, whose position
is closest to the event.\footnote{This approach of handling the overlap is different from the
equivalent procedure in H05. The latter tries to collect relatively {\it
pure} events, free of contamination from neighbors, but in the
process, it drops
valid source events if it is highly ambiguous which source they belong to.
The new, simple approach counts in all the events in the source
regions, which is more appropriate for estimating the total resolved
fraction later, although photometry results may suffer mild contamination from
neighbors.}  The aperture ratio of the source to background regions is adjusted
accordingly.


\subsection{Instrumental Background} \label{s:instbkg}

In aperture photometry of discrete sources, background subtraction
handles both the instrumental and diffuse X-ray background simultaneously.  
In order to calculate the total resolved fraction, one also has to know the
total incoming X-ray flux in the region, which requires an estimate
of the total instrumental background in the region.
As of this writing, two
stowed data sets (Period D and E) are available for modeling the instrumental 
background of the \chandra/ACIS
instruments.\footnote{http://cxc.cfa.harvard.edu/contrib/maxim/acisbg}
Period E is from 2005, Oct, 1 to the end of 2009, and Period D is from
2000, Dec, 1 to 2005, Aug, 31.  Therefore, the stowed data in Period
E is more appropriate for modeling the instrumental background of the
observations of the LW. The instrumental background summed over the
entire chip is shown
to be consistent
over the years \citep{Hicox06}. However, in HRES, both the count rate
and the spectral shape of the instrumental background show a noticeable
change between Period D and E, which is significant enough to change
the overall resolved fraction greatly (Appendix~\ref{s:inst}).
For instance, if Period D stowed data set is used alone, the total
resolved fraction of the iron line flux becomes more than 100\%,
which is in part due to the low statistics of the stowed data set in HRES.
We use the stowed data set of Period E alone and the combined set of
Period D and E for analysis.

\section{Results} \label{s:results}

Here we summarize the results of our analysis in comparison with R09.
As we explore several analysis approaches, the results are somewhat
extensive. However, they are more or less consistent with a few noticeable
exceptions, so we describe the main results based on the
default choice of the analysis parameters and point out any significant
variations resulted by other parameter choices.  The default choice
includes the EDSER procedure, the \VF mode cleaning for the 2008 observations,
and the boresight offset correction as
listed in Table~\ref{t:analysis}. 

\begin{figure*} \begin{center}
\noindent
\includegraphics*[width=0.99\textwidth]{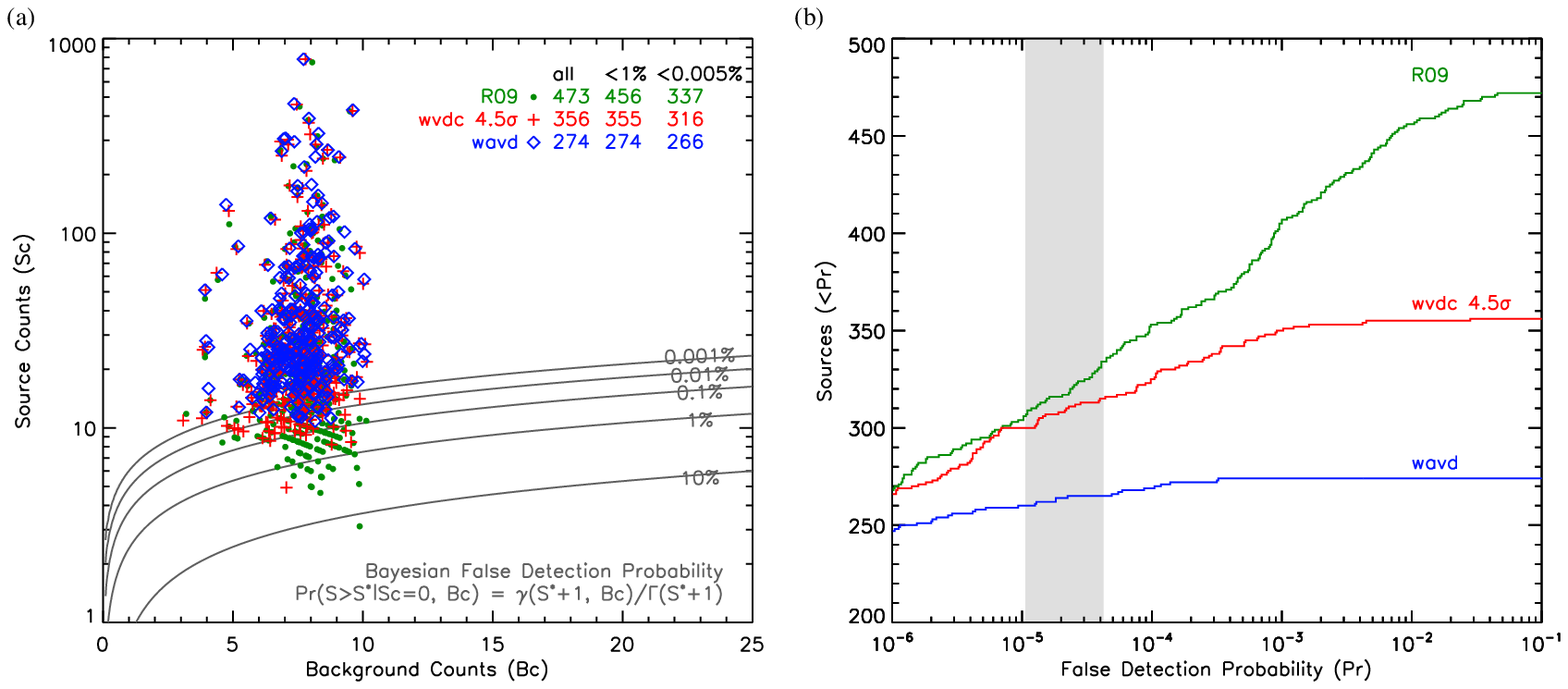}
\caption{(a) Distribution of 
the source and background counts in 
detection cells (a 1\arcsec\ radius circle around each source), overplotted
with false detection probability ($P_r$) calculated by the Bayesian analysis
\citep{Weisskopf07, Kashyap10}. See also Appendix~\ref{s:detsig}. (b)
Culumative distributions
of the source numbers as a funcion of $P_r$.
The shaded region indicates the limit required to ensure
1 or less false source in the source list, assuming that
24k to 96k trials were performed.
}
\label{f:detsigsrc}
\end{center}
\end{figure*}

\subsection{Source Detection} \label{s:detection}

Table~\ref{t:detection} summarizes the source search results under
the default parameter choice. 
Two search routines with three parameter
choices were applied to the 0.5--7 keV X-ray image of the field 
(column 2) with the exposure map generated at 1.5 keV. These results are
compared with the source number from R09.  We detected 251--274 sources from
the \wavdetect routine, depending on
how we processed the event files (Table~\ref{t:analysis}).  Therefore,
R09 detected 70--90\% more sources than what the \wavdetect routine
discovered. Compared to our search using the \wvdecomp routine under
the presumably similar parameter settings (4.0$\sigma$) as in R09,
R09 still found about 2--6\% more sources.
Figure~\ref{f:image} shows three of the four search results.  In order
to evaluate significance of these detections, we take both 
programmatic and analytic approaches.

\subsubsection{Programmatic Approaches}

First, we applied the same search routines to the 0.5--7 keV image
generated from the stowed data that were reprojected to the sky according
to the aspect solution of the observations (column 10). For the analysis
with the \VF mode cleaning, only the events with flag=0 were reprojected
in simulating the portion of the 2008 observations.  For no \VF mode
cleaning, all the stowed data
were reprojected (not shown in Table~\ref{t:detection}).

The reprojected stowed data can provide a good indicator of false
detection rate since they are based on the actual events and the proper
dithering motion is included through the reprojection, but there are a
few short comings. For instance, the equivalent exposure of the stowed
data in Period E is 360 ks, which is a factor of three shorter
than the LW data set to reflect the proper Poisson fluctuation for the 1
Ms observation.  In addition, events in ACIS-I CCD 1 of the stowed data
are artificially generated, based on events in CCD 0.  We also note the
X-ray flux in the region below $\sim$ 5 keV remained largely
unresolved (\S\ref{s:resolved}, see also R09).  Since the image
used for source search is generated in the 0.5--7 keV band, 
the input image for source search routines contains a large diffuse
(or unresolved) component besides the instrumental background, which
can enhance false detection rate, but the effect of this diffuse sky
component cannot be properly accounted for with the stowed data set.
Therefore, the source number from the reprojected stowed data set
(e.g.~25 in the 439 sources for \wvdecomp (4.0$\sigma$)
in Table~\ref{t:detection}) represents only a lower limit of the false
detections.

The legitimacy of the 473 sources in R09 is in part based on their claim
that the same search routine found only 1 or 2 (false) sources on the
stowed data set.  The stowed data set they used is likely an earlier
version of the ones we use and each reprojection generates a different
data set in the sky, so there can be some fluctuations from run to run.
However according to our analysis, 1 or 2 false detections in the 473
sources of R09 appear to be a severe underestimate.

Second, we applied the search routines to the 9--12 keV image of the LW
data (column 9), where no discrete sources are expected to be discovered
due to the diminishing effective area
($\lesssim$10 cm\sS{2}) and the large PSF.  This provides another estimate
of the false detections (e.g.~24 in the 439 sources for \wvdecomp (4.0$\sigma$) in
Table~\ref{t:detection}), but they are still lower limits since the
9--12 keV image also have the similar shortcomings as the stowed data set
(e.g. the total number of events in the 9--12 keV band is much smaller
than that in the 0.5--7 keV band).

Therefore, we took one more approach to address the false detection rate.
We compared the four source lists to find out the objects that
are not common to the lists as shown in Table~\ref{t:detection}
(Columns 5--8).\footnote{
For source match, we employed the prescription 
by \citet{Brandt01}; M09 with a bit looser requirement by allowing a
0.5\arcsec\ additional offset to compensate for a possible binning effect
in the image and source search; i.e. in HRES, when the source separation
is more than 1.1\arcsec, then they are considered to be different.
Therefore, the sources unique to each list
Table~\ref{t:detection} are usually located in a different section of
the HRES region as seen in Figure~\ref{f:image}, and they are indeed different
sources (not misrecognized as unique to a list because of the pixellation or
any other similar effects).
}
Detection of faint sources near the detection limit is expectedly 
sensitive to small changes in the image. For instance, simply changing
the image pixellation offset (e.g.~2901.5:5416.5:\#2515
vs.~2901.0:5416.0:\#2515) or the reprojection tangential point (Column
8) produces a different set of sources under the otherwise identical
procedures (i.e.~42 different sources for \wvdecomp with 4.0$\sigma$
from the tangential point change). Obviously one cannot rule out all
of these list-unique sources as invalid, but it is clear that they are
prone to small statistical fluctuations of the image and less reliable
than the sources consistently detected through these variations. 
While our source number count (439) did not reach that by R09 (473)
under a similar \wvdecomp run, the comparison of our source list with
the list by R09 indicates that about 80--140 sources are in fact
unique to each list.

We find a large number of the sources that are not common in our search and R09
in Table~\ref{t:detection} puzzling. Are all of  these sources,
which now add up to $\sim$ 550 objects, valid?
To address this, we turn to analytic approaches.  

\begin{table*}
\footnotesize
\caption{Aperture Photometry Results of the Resolved Discrete Sources in HRES
in the 6.5 -- 7.1 and 9 -- 12 keV bands}
\begin{tabular*}{\textwidth}{l@{\extracolsep{\fill}}D{(}{\ (}{5.3}D{(}{\ (}{5.3}D{(}{\ (}{5.3}D{(}{\ (}{5.3}}
\hline
\multicolumn{1}{c}{Options / Energy Band}                  & \multicolumn{1}{c}{wavd}        & \multicolumn{1}{c}{wvdc ($4.5\sigma$)} & \multicolumn{1}{c}{wvdc ($4.0\sigma$)} & \multicolumn{1}{c}{R09}         \\
\hline
1) Default                                                 & \multicolumn{1}{c}{274 sources} & \multicolumn{1}{c}{356 sources}        & \multicolumn{1}{c}{439 sources}        & \multicolumn{1}{c}{473 sources} \\
\multicolumn{1}{r}{Total Events in 6.5--7.1 keV}           & 494(22)                         & 569(24)                                & 636(25)                                & 690(26)                         \\
\multicolumn{1}{r}{9.0--12. keV}                           & 1642(41)                        & 2099(46)                               & 2560(51)                               & 2818(53)                        \\
\cline{2-5}\multicolumn{1}{r}{Net Photons in 6.5--7.1 keV} & 295(22)                         & 316(24)                                & 328(26)                                & 355(27)                         \\
\multicolumn{1}{r}{9.0--12. keV}                           & 23(41)                          & 32(47)                                 & 34(53)                                 & 55(55)                          \\
\hline 2) 1.5 keV 95\% PSF with Annulus Bkg.               & \multicolumn{1}{c}{274 sources} & \multicolumn{1}{c}{356 sources}        & \multicolumn{1}{c}{439 sources}        & \multicolumn{1}{c}{473 sources} \\
\multicolumn{1}{r}{Net Photons in 6.5--7.1 keV}            & 303(23)                         & 322(24)                                & 325(25)                                & 355(27)                         \\
\multicolumn{1}{r}{9.0--12. keV}                           & 17(41)                          & 0(47)                                  & 2(52)                                  & 1(54)                           \\
\hline 3) \F mode                                          & \multicolumn{1}{c}{264 sources} & \multicolumn{1}{c}{354 sources}        & \multicolumn{1}{c}{458 sources}        & \multicolumn{1}{c}{473 sources} \\
\multicolumn{1}{r}{Net Photons in 6.5--7.1 keV}            & 308(25)                         & 322(27)                                & 349(30)                                & 366(30)                         \\
\multicolumn{1}{r}{9.0--12. keV}                           & -3(53)                          & 28(61)                                 & 19(70)                                 & 65(72)                          \\
\hline 4) No Boresight                                     & \multicolumn{1}{c}{267 sources} & \multicolumn{1}{c}{351 sources}        & \multicolumn{1}{c}{446 sources}        & \multicolumn{1}{c}{473 sources} \\
\multicolumn{1}{r}{Net Photons in 6.5--7.1 keV}            & 293(22)                         & 295(24)                                & 315(26)                                & 342(27)                         \\
\multicolumn{1}{r}{9.0--12. keV}                           & 10(41)                          & 33(47)                                 & 77(53)                                 & 31(55)                          \\
\hline
\end{tabular*}

\label{t:ap}
\end{table*}

\subsubsection{Analytic Approaches} \label{s:detanal}

In order to estimate detection significance ($C$) or false detection
probability ($P_r=1-C$) {\it independent} of the source search routines,
we employ
a Bayesian approach by \citet{Weisskopf07, Kashyap10}, which provide a
more rigorous treatment of discrete Poisson distributions than simple
minded approaches using signal-to-noise ratio (SNR).  In
Appendix~\ref{s:detsig}, we describe a simplified version of the Bayesian
approach used for calculating the detection significance and compare it
with the SNR based analysis.

Columns 3 and 4 in Table~\ref{t:detection} show the number of sources
with $P_r$$<$1\% and $<$0.005\% respectively.  The results in Column
3 give a false impression that the majority of these sources are
significant.  When dealing with a single source or a known source in
a new observation, finding the source with a confidence level at 99\%
($P_r$=1\%) can be considered sufficient to claim a true detection.
However, in studying a population of sources newly discovered by a
search algorithm, one has to consider the number of search trials
explicitly, e.g.~99\%
confidence means 1 out of 100 {\it trials} can be false. 

The source search routines conduct searches in the
entire input images using a small window or detection cell.  The cell size
is usually smaller than the source aperture region
used in aperture photometry for efficient source detection. 
Following the description of \citet{Weisskopf07}, we assume that a
1\arcsec\ radius circle is appropriate for the cell size for \chandra
images especially in the central regions like HRES.  The number of
independent search trials can be roughly estimated as the ratio of the
search region to the cell size.  In the HRES region, about 
24k independent search trials can be performed.  Under 2-Dim Nyquist sampling,
these numbers quadruple (96k).

For simplicity we assume each search routine performed 20k
trials in HRES,\footnote{This is a conservative estimate.  Our main
point, the need for high detection threshold, remains valid as long as
the search trials $\gg$ 1.  The precise trial number depends on each
algorithm. e.g.~\wvdecomp employs iterations for finding faint sources,
so the actual number can be much larger than the ratio of the search
region to the cell size.  The \wavdetect routine internally considers
each pixel as an independent trial, and the default threshold at $10^{-6}$
means allowing one false source in 1 Mpixels (1 ACIS CCD).  We believe
that the detection cell size, which is proportional to the PSF size,
should be accounted for in order to get the truly independent trial
numbers.  However, considering that the two closest sources in the 473
sources by R09 are about a pixel apart, the trial statistics counting
scheme in \wavdetect may be also appropriate for the \wvdecomp run by
R09, in which case there are about 300k trials in HRES.} then having a
detection with 99\% confidence means that there can be as many as 200
false detections arising from random Poisson fluctuations.  Note that
not all of these 200 sources will be in the source list since each search
routine has its own selection criteria to remove false sources.  What this
means is that sources with 99\% confidence have the same significance
of other 200 false sources that can be found in the search region.
Therefore, in order to make sure the source list contains 1 or less
false sources, the required confidence level should be 99.995\% or higher
(Column 4).

Figure~\ref{f:detsigsrc} shows the distribution of source and background
counts (see Appendix~\ref{s:detsig}) in detection cell (a 1\arcsec\ radius circle
around each source), which is overplotted with various levels of $P_r$.
The figure also shows the culumative distributions of the sources as a function
of $P_r$.  The shaded region indicates the limit required to ensure 1
or less false source in the source list, corresponding to 24--96k
trials of source search. Figure \ref{f:detsigsrc} indicates
that $\lesssim$ 337 out of the 473 sources in R09 are detected with
sufficiently high significance.

These analytic approaches also often provide only a lower limit of false
detection rates due to missing implementation of (usually unknown) subtle
features in real data that can give rise to false detections (e.g. source
crowding, node boundaries of CCDs or dithering motion induced event
scattering).  Therefore, we believe the need for detection confidence
level more stringent than 99\% for source selection remains valid.

\begin{figure*} \begin{center}
\noindent
\includegraphics*[width=0.99\textwidth]{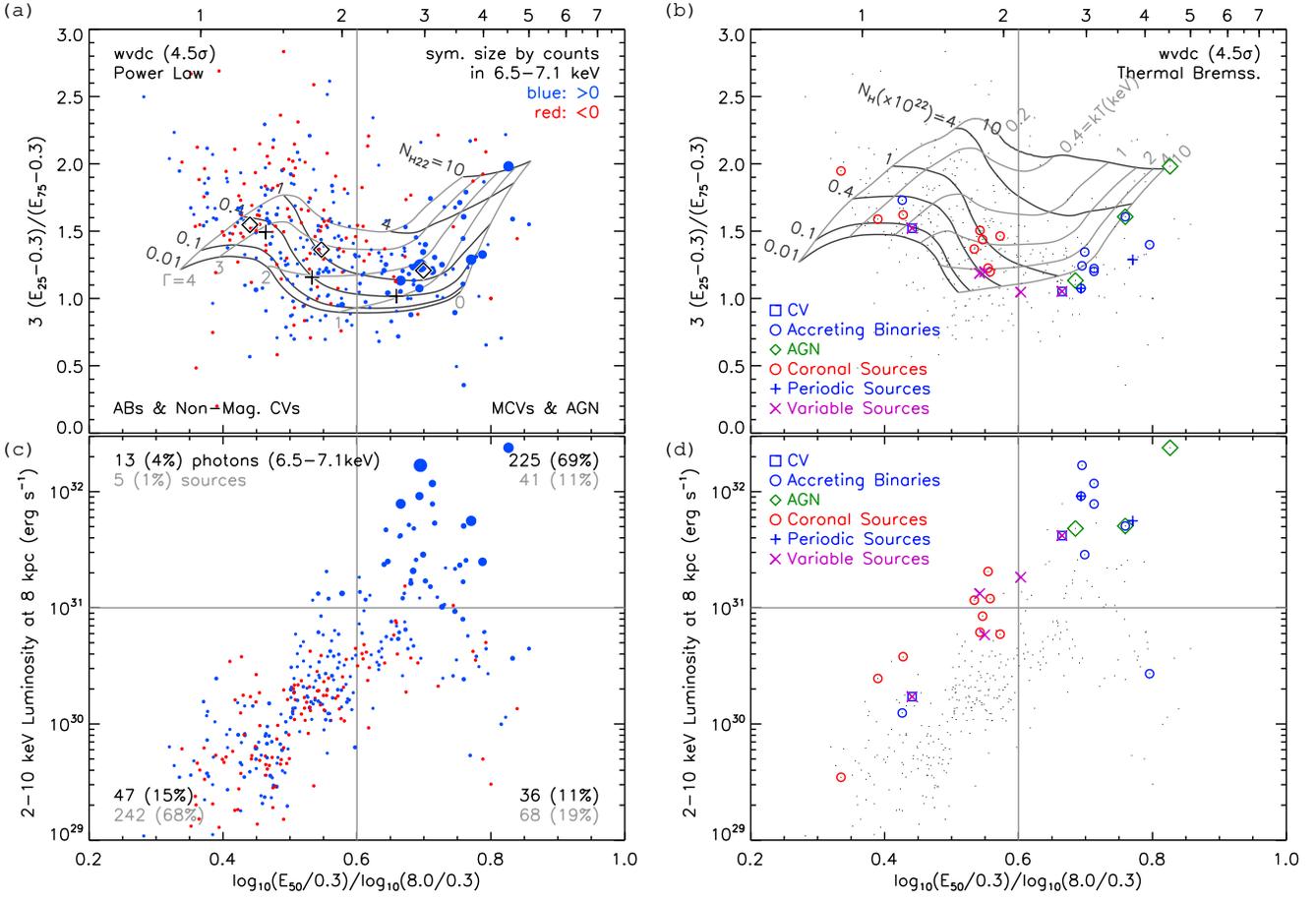}
\caption{X-ray spectral and luminosity distributions of the 356
sources found by the \wvdecomp (4.5$\sigma$) routine in HRES of the LW. 
(a) Quantile diagram \citep{Hong04,Hong09a} overlayed with the power
law model grids, and the
symbol size is semi-logarithmically proportional to the absolute net
counts in the 6.5--7.1 keV band.
The blue and
red dots represent positive and negative net counts respectively.
The diamond and cross symbols represent the combined X-ray spectra
of the groups selected by the median values (Q1, Q2 \& Q3 in
Table~\ref{t:conv}) and net counts (N1, N2 \& N3) respectively.
(b) the same as (a) with the thermal Bremsstrahlung model grids, and two dozen sources are marked
according to their likely type. (c) \& (d) The median energy vs.~the
2--10 keV X-ray luminosity at 8 kpc. In (c), the fractional
contribution of source number (grey) and the 6.5--7.1 keV net photon
counts (black) are noted in each quadrant of the diagram.
}
\label{f:qtil}
\end{center}
\end{figure*}

\begin{table*}
\begin{minipage}{0.99\textwidth}
\caption{Count to Flux Conversion Factor and X-ray Luminosity by Source Group}
\begin{tabular*}{\textwidth}{lccrD{(}{(}{1.1}@{\extracolsep{\fill}}D{(}{(}{3.1}D{.}{.}{1.4}D{.}{.}{1.4}D{.}{.}{1.4}D{.}{.}{2.1}D{.}{.}{1.2}D{.}{.}{1.2}D{.}{.}{2.1}}
\hline
\multicolumn{1}{c}{(1)}  & \multicolumn{1}{c}{(2)}       & \multicolumn{1}{c}{(3)} & \multicolumn{1}{c}{(4)}    & \multicolumn{1}{c}{(5)}               & \multicolumn{1}{c}{(6)}        & \multicolumn{1}{c}{(7)}        & \multicolumn{1}{c}{(8)}  & \multicolumn{1}{c}{(9)}             & \multicolumn{1}{c}{(10)}         & \multicolumn{1}{c}{(11)}        \\
                         & \multicolumn{2}{c}{Selection}                        & \multicolumn{1}{c}{Source} & \multicolumn{2}{c}{Net Photon Counts}                               &                                &                          &                                     & \multicolumn{1}{c}{Unabs.}       & \multicolumn{1}{c}{$L_X$ at}    \\
\cline{2-3}\cline{5-6}ID & Net                           & QDx                     & \multicolumn{1}{c}{Number} & \multicolumn{1}{c}{6.5--7.1}          & \multicolumn{1}{c}{0.5--7 keV} & \multicolumn{1}{c}{$E$\Ss{50}} & \multicolumn{1}{c}{\nHt} & \multicolumn{1}{c}{$\Gamma$}        & \multicolumn{1}{c}{Flux}         & \multicolumn{1}{c}{8 kpc}       \\
                         & Counts                        &                         &                            &                                       &                                & \multicolumn{1}{c}{(keV)}      &                          &                                     & \multicolumn{1}{c}{($10^{-17}$)} & \multicolumn{1}{c}{($10^{29}$)} \\
\hline
Q3                       & $\le$300                      & $<$0.5                  & 123                        & 19(10)                                & 4181(89)                       & 1.27(1)                        & 0.31(5)                  & 3.4(2)                              & 0.29                             & 0.22                            \\
Q2                       & $\le$300                      & 0.5..0.6                & 121                        & 37(11)                                & 4098(88)                       & 1.81(2)                        & 1.11(8)                  & 2.8(1)                              & 1.20                             & 0.91                            \\
Q1                       & $\le$300                      & $\ge$0.6                & 102                        & 171(16)                               & 5628(94)                       & 2.98(4)                        & 1.5(1)                   & 1.6(1)                              & 2.87                             & 2.19                            \\
\hline
B                        & $>$300                        &                         & 10                         & 95(10)                                & 4364(68)                       & 2.65(4)                        & 0.24(7)                  & 0.86(6)                             & 2.59                             & 1.97                            \\
\hline
N1                       & $>$100                        &                         & 35                         & 203(15)                               & 8583(98)                       & 2.61(3)                        & 0.40(5)                  & 1.04(6)                             & 2.50                             & 1.91                            \\
N2                       & 10..100                       &                         & 293                        & 125(19)                               & 9529(137)                      & 1.73(1)                        & 0.37(6)                  & 1.94(8)                             & 1.23                             & 0.94                            \\
N3                       & 5..10                         &                         & 19                         & 0(5)                                  & 142(26)                        & 1.4(1)                         & 0.3(7)                   & \multicolumn{1}{D{(}{(}{1.4}}{3(2)} & 0.49                             & 0.37                            \\
\hline
\end{tabular*}

Notes. Based on the 356 sources detected by \wvdecomp (4.5$\sigma$). 
(1) Group IDs. (3) QDx = $\log(E_{50}/0.3)/\log(8/0.3)$. It is a normalized
logarithmic value of the median energies, $x$-axis values of the quantile diagram \citep{Hong09a}.
(5) \& (6) Summed net photon counts of the sources in the group.
(7) The median energy of the combined X-ray spectrum of the group.
(8) \& (9) Interstellar absorption and power law index of the combined
X-ray spectrum of the group under a simple power law model.
(10) The unabsorbed 2--10 keV flux in $10^{-17}$ \fcgs for sources with
1 net photon in the 0.5--7 keV band or Conversion factor from the
0.5--7 keV counts to the 2--10 keV unabsorbed X-ray flux in $10^{-17}$
\fcgs ph\sS{-1}. 
(11) The 2--10 keV luminosity at a distance of 8 kpc in $10^{29}$ \lcgs
for sources with 1 net photon in the 0.5--7 keV band. 
\label{t:conv}
\end{minipage}
\end{table*}

In summary, the results in Table~\ref{t:detection} indicate that the
faintest $\sim$ 100 sources in R09 may not be as significant as R09
claimed.  We wonder whether the
search parameters in R09 may have been pushed beyond the reasonable
limit.  On the other hand, our results directly conflict with the
aperture photometry results of R09. Figure~3 in R09 indicates a large
contribution in the resolved flux in the 6.5--7.1 keV band comes from
the faintest sources, which is difficult to imagine if they are mostly
false or insignificant detection.  We will explore this through aperture
photometry in the next section.

\begin{figure*} \begin{center}
\includegraphics*[width=0.99\textwidth]{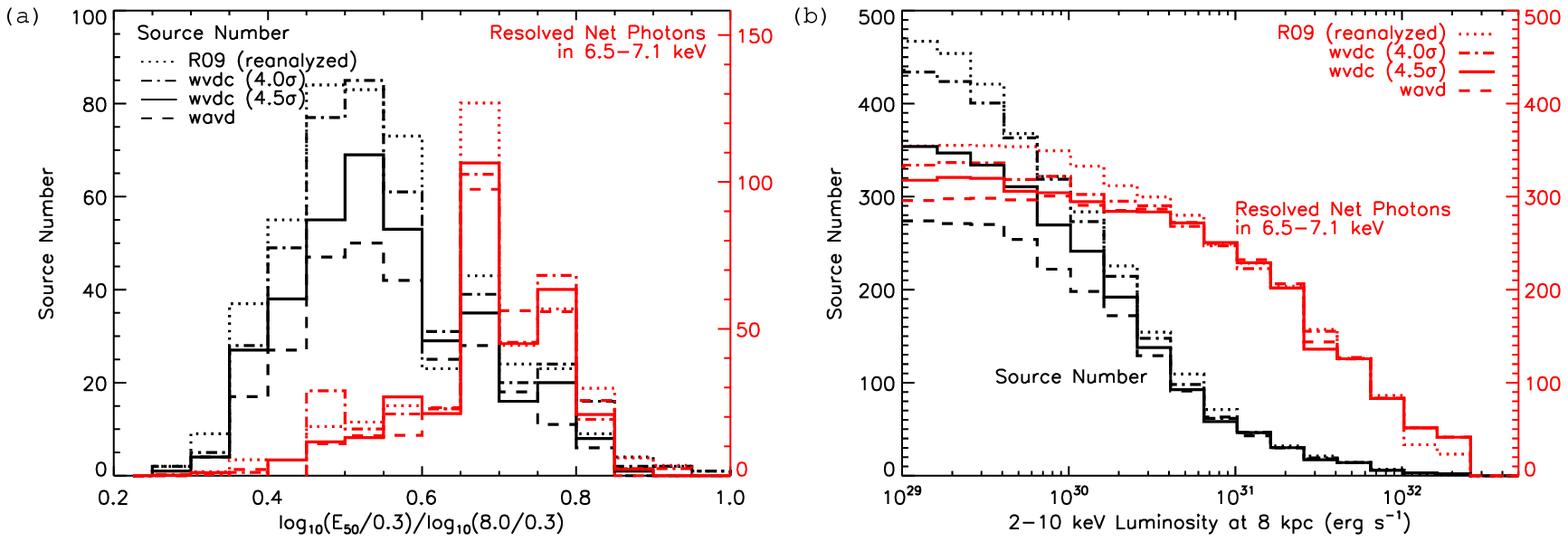}
\caption{X-ray spectral and luminosity histograms of the resolved sources in HRES. 
(a) Comparison of the source number (black) and 6.5--7.1 keV net count (red) distributions
as a function of median energies ($E\Ss{50}$). (b) same as (a), but the cumulative distributions 
as a function of the 2--10 keV X-ray luminosity at 8 kpc. The results of
three different source searches are shown: the 274 sources by \wavdetect (dashed),
the 356 sources by \wvdecomp (4.5$\sigma$, solid), and the 473 sources
by R09 (dotted). The majority of the resolved net photons are from
the brightest 100-150 sources.
Note the faintest ends
of the cumulative net count distributions in the 6.5--7.1 keV band
alternate positive and negative slopes due to many faint sources with negative net counts in the band.}
\label{f:dist}
\end{center}
\end{figure*}

\subsection{Aperture Photometry} \label{s:ap_results}

Table~\ref{t:ap} summarizes the aperture photometry results of the resolved
discrete sources in the HRES region. The table shows the summed total
events in the source aperture regions and net photon counts after background
subtraction in the 6.5--7.1 and 9--12 keV bands for the various analysis
options.  The default choice uses a fixed 2\arcsec\ radius aperture
around each source for the source region and the rest of the HRES regions
(excluding other source apertures) for the background region; the data
were prepared and merged with the EDSER procedure, the \VF mode
cleaning and the boresight offset correction.  The table compares two
aperture photometry methods (Option 1 vs.~2), event selection methods
(Option 1 vs.~3), and the effects of the boresight offset correction
(Option 1 vs.~4).  For the default choice, we also show the total
events in the source region before the background subtraction.

The 6.5--7.1 keV band is chosen to represent the emission lines from
highly ionized irons as in R09.  The 9--12 keV band results are shown
for sanity check: despite the large number of the total events in the
source regions in the 9--12 keV band (1642--2818), which are roughly
proportional to the number of sources, the summed net photon counts
after background subtraction are essentially null, consistent with random
Poisson fluctuations as expected.

The four analysis options in Table~\ref{t:ap} produce essentially
identical results, indicating the outcomes of the aperture photometry
are very robust.  Only the case with no boresight offset correction
(Option 4) produces consistently lower net photon counts for all the four
source search routines. While the differences are still within the
statistical fluctuations from the other results, the consistent deficit
in the net photon counts with no boresight offset correction implies
that the boresight offset correction was applied properly in the other
options and does improve the aperture photometry.

Unlike the total event counts which are roughly proportional to the
number of sources, the total 6.5--7.1 keV net photon counts after
background subtraction do not vary significantly among the four source lists.
The net photon counts are consistent within $\sim$ 3$\sigma$, despite
the large differences in the source numbers (e.g.~in Option 1, 295 $\pm$
22 photons from the 274 sources by \wavdetect vs.~355 $\pm$ 27 photons from
the 473 sources by R09). The source lists by the three \wvdecomp routines
produce the consistent results within $\sim$ 2 $\sigma$, indicating there
is no significant contribution from additional $\sim$ 100 sources found
by R09 in comparison to the 356 sources found by the \wvdecomp algorithm
with 4.5$\sigma$.  These aperture photometry results are consistent
with the conclusion of the source search results in \S\ref{s:detection}:
the 50--100 faintest sources in R09 do not contribute significantly in the
resolved flux in 6.5--7.1 keV band.  This is very different from
the conclusion in R09.  In order to find the origins of the discrepancy,
we explore the spectral and luminosity properties of these sources in more
detail.

\subsection{Spectral and Luminosity Distribution of the Resolved X-ray Sources} \label{s:ap_res2}

Figure \ref{f:qtil} shows the spectral and X-ray luminosity distributions of the 356 sources
detected by the \wvdecomp routine (4.5$\sigma$).   These sources
are chosen for illustration, and the conclusion remains unchanged
for other source search results (see Figure~\ref{f:dist}).  
Panels (a) \& (b) plot the energy
quantiles \citep{Hong04,Hong09a} of the sources in the same phase space with two spectral model
grids (powerlaw and thermal Bremsstrahlung) for comparison.  Panels (c) \&
(d) display the same sources in a phase space of the median energy
($E\Ss{50}$, in the 0.3--8 keV band) and the
2--10 keV X-ray luminosity (see below) at a distance of 8 kpc (the Galactic Center).
The symbol sizes in (a)  \& (c) are semi-logarithmically proportional to
the absolute counts of net photons in the 6.5--7.1 keV band. The blue and
red dots represent positive and negative net counts respectively. Panels
(b) \& (d) mark a few dozen identified sources or
sources with some clue about their nature \citep[V09, H12,][]{Hong12b}.

\subsubsection{Diverse Spectral Types and Flux Calculation}

Figure~\ref{f:qtil} illustrates the diverse spectral types
of X-ray sources in the region and the results are very intuitive:
the bright, hard X-ray sources mostly contribute the hard X-ray flux. 
Table~\ref{t:conv} also shows the spectral diversity by grouping similar
sources. In order to estimate the flux and luminosity of each source in
the 2--10 keV band, R09 relies solely on the
net counts in the 0.5--7 keV net count rate (see Appendix~\ref{s:conv}). 
Both Figure~\ref{f:qtil} and Table~\ref{t:conv} do show a strong
correlation between the spectral type and the net counts.  However,
the spectral diversity present in the same count range of the sources
indicates that the count-to-flux conversion factor only based on the
counts underestimates the spectral diversity and misassign the flux
values of the sources.  Therefore, we group them by the median energy
of the sources (only for the sources with less than 300 counts, where
the spectral model fit for each source is not reliable).  In this way,
each group has more or less similar spectral types of sources, and the
conversion factor from count rate to flux will reflect their spectral
properties. Table~\ref{t:conv} shows about a factor of 10 variation in
the conversion factor between the softest (Q3) and hardest groups (Q1).
It also shows the combined spectrum of the brightest sources (B) is most
consistent with the hardest sources (Q1).
The 2--10 keV X-ray luminosity at 8 kpc in Figure~\ref{f:qtil} are
calculated using the conversion factors from the net photon counts in the
0.5--7 keV band of three median energies based groups under a simple power
law model.  For the bright sources with $\ge$ 300 net photons, we use the
result of a spectral model fit to each spectrum using a power law model.
We have also tried thermal plasma models such as thermal Bremsstrahlung,
and for those that produce reasonable fits to the spectra, the X-ray
luminosity estimates come out consistent with the one from the power law
model.

\subsubsection{Sources with Clues}

Two best candidate CVs are classified as such based on the blue color, \Ha
excess of the optical counterpart, and high $F_X/F_R$ ratio.
Accreting binaries or more marginal candidate CVs are based on the
blue color and high $F_X/F_R$ ratio. Coronal
sources are the OGLE-III\footnote{\url{http://ogle.astrouw.edu.pl}}
variables or have bright
UCAC2\footnote{\url{http://ad.usno.navy.mil/ucac/u2_readme.html}}
counterparts (see V09 and references therein).
Three (candidate) AGN are based on their hard X-ray spectrum and high
absorption: one with an extended optical counterpart, 
another with a very red counterpart and the other with
a blue counterpart (i.e.~it can also be an accreting binary). 
These sources are based on the inital 100
ks \chandra ACIS-I observations of the LW (V09), and a
similar study using the sources from the full 1 Ms exposure is underway
\citep{Berg12b}.  Two periodic sources found in the region are likely MCVs (H12),
one of which is recognized as a candidate accreting binary in
(V09).  Two of four non-periodic variable sources (flaring
or transient)
found in the region \citep{Hong12b}
are also the two best candidate CVs in V09.
Periodic or variable X-ray sources are identified from the 
1 Ms exposure data.

In the LW outside of the HRES region, a few dozen more sources are either
identified or show some clues about their nature (H12, V09). 
Their distribution shows a similar pattern in the quantile diagram, namely
MCVs and AGN are dominantly located at median energy 
$E\Ss{50}$ $\gtrsim 2.2$ keV, whereas
non-magnetic CVs and coronal sources such as ABs are at $E\Ss{50}$ $\lesssim 2.2$ keV
as illustrated by the vertical gray line in the diagram (see V09, H12).  
For instance, Figure 9 in H12 shows all 10 MCVs identified
in the LW through their periodic X-ray modulation are at $E\Ss{50}$ $\gtrsim 2.2$ keV.
The symbol
size and color clearly show the large contribution to the 6.5-7.1
keV flux from the sources located at $E\Ss{50}$ $\gtrsim 2.2$ keV despite their relative
paucity in the diagram (see also Figure~\ref{f:dist}).  In the case of the luminosity distribution,
the dominant contribution to the 6.5-7.1 keV flux comes from the relatively bright
sources ($\gtrsim$$10^{31}$ \lcgs).  Panel (c) in figure~\ref{f:qtil} shows the source fraction and the 6.5-7.1 keV net
count contribution of each quadrant in the diagram. The bright,
hard sources contribute about 70\% of the  6.5-7.1 keV net counts,
although they are only about 10\% of the total source number.

\subsubsection{Sources Resolving Iron Line Flux}
 
Figure~\ref{f:dist} shows the spectral and luminosity distributions
of the sources found in HRES. Panel (a) contrasts the source number
distribution (black lines), which is dominated by the soft sources, with
the resolved 6.5--7.1 keV net counts (red lines), which are dominated
by the hard sources.  The same trends are visible in different source
lists by \wavdetect (dashed lines), \wvdecomp (4.5$\sigma$, solid)
and R09 (dotted). Panel (b) shows the cumulative source number and
6.5--7.1 keV net counts as a function of the 2--10 keV X-ray luminosity
for three source lists.  The source number distribution (black lines)
are again distinct from the resolved 6.5--7.1 keV net count distribution
(red lines).  Although comprising a relatively small fraction of the total
source number, the relatively bright and hard X-ray sources contribute
most of the resolved iron line flux.  The small increase in the resolved
fraction as the source number counts increases from 274 by \wavdetect
to 473 by R09 can be in part due to a small addition of real sources
as expected from the lowered detection threshold
\citep[false negatives, see][]{Kashyap10}.  
We will address the unresolved spectra in the next section
(e.g.~Figure~\ref{f:breakup})

\begin{table}
\footnotesize
\caption{Photon Count in HRES in 3--7, 6.5--7.1 and 9--12 keV}
\begin{tabular*}{0.47\textwidth}{l@{\extracolsep{\fill}}rD{(}{\ (}{5.3}D{(}{\ (}{5.3}}
\hline\hline
\multicolumn{1}{c}{Band}    & \multicolumn{1}{c}{Data} & \multicolumn{1}{c}{Stow E} & \multicolumn{1}{c}{Stow DE} \\
\multicolumn{1}{c}{Options} & \multicolumn{1}{c}{}     & \multicolumn{1}{c}{}       & \multicolumn{1}{c}{}        \\
\hline
3--7 keV                    & All the Events           & \multicolumn{2}{c}{38412(196)}                             \\
\VF clean                   & Reproj.~Stowed           & 10196(91)                  & 14650(109)                  \\
                            & Net Photon Counts        & 11787(397)                 & 12257(352)                  \\
                            & \sS{*}Surface Brightness & 492(17)                    & 514(15)                     \\
\hline 6.5--7.1 keV         & All the Events           & \multicolumn{2}{c}{4520(67)}                             \\
\VF clean                   & Reproj.~Stowed           & 1531(35)                   & 2271(43)                    \\
                            & Net Photon Counts        & 523(119)                   & 465(107)                    \\
                            & \sS{*}Surface Brightness & 77(18)                     & 68(16)                      \\
\hline 9--12 keV            & All the Events           & \multicolumn{2}{c}{34424(186)}                             \\
\VF clean                   & Reproj.~Stowed           & 13183(102)                 & 19282(123)                  \\
                            & Obs.~to Stowed Ratio     & 2.61(2)                    & 1.79(1)                     \\
\hline\hline 3--7 keV       & All the Events           & \multicolumn{2}{c}{45962(214)}                             \\
\F mode                     & Reproj.~Stowed           & 12663(113)                 & 18206(135)                  \\
                            & Net Photon Counts        & 11948(459)                 & 12169(406)                  \\
                            & \sS{*}Surface Brightness & 508(20)                    & 513(17)                     \\
\hline 6.5--7.1 keV         & All the Events           & \multicolumn{2}{c}{6698(82)}                             \\
\F mode                     & Reproj.~Stowed           & 2254(47)                   & 3328(58)                    \\
                            & Net Photon Counts        & 643(159)                   & 520(142)                    \\
                            & \sS{*}Surface Brightness & 95(23)                     & 77(21)                      \\
\hline 9--12 keV            & All the Events           & \multicolumn{2}{c}{58209(241)}                             \\
\F mode                     & Reproj.~Stowed           & 21670(147)                 & 31360(177)                  \\
                            & Obs.~to Stowed Ratio     & 2.69(2)                    & 1.86(1)                     \\
\hline
\end{tabular*}

\sS{*}Surface Brightness: 10\sS{-13} \fcgs deg\sS{-2}
\label{t:count}
\end{table}

\begin{table*}
\footnotesize
\begin{minipage}{0.99\textwidth}
\caption{Resolved Fraction in HRES in 6.5 -- 7.1 keV}
\begin{tabular*}{\textwidth}{l@{\extracolsep{\fill}}D{+}{\pm}{5.3}D{+}{\pm}{5.3}D{+}{\pm}{5.3}D{+}{\pm}{5.3}}
\hline
Fix 2\arcsec\ Radius Aperture                    & \multicolumn{1}{c}{\wavdetect}    & \multicolumn{1}{c}{\wvdecomp (4.5$\sigma$)} & \multicolumn{1}{c}{\wvdecomp (4.0$\sigma$)} & \multicolumn{1}{c}{\wvdecomp (R09)} \\
\hline \VF mode clean                            & \multicolumn{1}{c}{(274 sources)} & \multicolumn{1}{c}{(356 sources)}           & \multicolumn{1}{c}{(439 sources)}           & \multicolumn{1}{c}{(473 sources)}   \\
\multicolumn{1}{r}{Source to Bkgnd Region Ratio} & \multicolumn{1}{c}{4.9\%}         & \multicolumn{1}{c}{6.4\%}                   & \multicolumn{1}{c}{7.9\%}                   & \multicolumn{1}{c}{8.7\%}           \\
\multicolumn{1}{r}{Resolved Fraction (Stow E)}   & 61 + 16 \mbox{\%}                 & 65 + 17 \mbox{\%}                           & 68 + 18 \mbox{\%}                           & 73 + 19 \mbox{\%}                   \\
\multicolumn{1}{r}{(Stow DE)}                    & 68 + 18 \mbox{\%}                 & 73 + 19 \mbox{\%}                           & 76 + 20 \mbox{\%}                           & 83 + 22 \mbox{\%}                   \\
\hline \F mode                                   & \multicolumn{1}{c}{(264 sources)} & \multicolumn{1}{c}{(354 sources)}           & \multicolumn{1}{c}{(458 sources)}           & \multicolumn{1}{c}{(473 sources)}   \\
\multicolumn{1}{r}{Source to Bkgnd Region Ratio} & \multicolumn{1}{c}{4.8\%}         & \multicolumn{1}{c}{6.4\%}                   & \multicolumn{1}{c}{8.3\%}                   & \multicolumn{1}{c}{8.7\%}           \\
\multicolumn{1}{r}{Resolved Fraction (Stow E)}   & 52 + 15 \mbox{\%}                 & 54 + 15 \mbox{\%}                           & 59 + 17 \mbox{\%}                           & 62 + 17 \mbox{\%}                   \\
\multicolumn{1}{r}{(Stow DE)}                    & 64 + 20 \mbox{\%}                 & 67 + 21 \mbox{\%}                           & 73 + 22 \mbox{\%}                           & 76 + 23 \mbox{\%}                   \\
\hline
\end{tabular*}

\label{t:resolved}
Notes. -- For the LW data, which were observed in 2005 and 2008, 
the Period E stowed data set is more appropriate (\S\ref{s:instbkg}). The
results using the Period D stowed data are shown just for comparison
to illustrate the dominance of the statistical fluctuation of the instrumental
background in the uncertainty of the resolved fraction. See Appendix~\ref{s:inst}.
\end{minipage}
\end{table*}

\subsection{Resolved Fraction of the Iron Emission Lines}  \label{s:resolved}

Table~\ref{t:count} summarizes the total events and net photon counts in
the HRES region in the 3--7, 6.5--7.1 and
9--12 keV bands.  The total net photons in the HRES region are calculated
by subtracting the instrumental background from the total event counts
in the region.  The instrumental background counts are acquired from
the reprojected stowed events in the same region. For subtraction,
we matched counts in the 9--12 keV band.\footnote{This is identical to the 
procedure in R09, although in R09 the count rate in the 9-12 keV band
is matched instead of the counts. It is because their subtraction is done
in count rate space and ours in count.} The total measured X-ray
surface brightness in HRES is $I$\Ss{3-7\ keV} = $(4.9 \pm 0.2)
\times 10$\sS{-11} \sbcgs, which is consistent with the result in R09 ($(4.6 \pm 0.4)
\times 10$\sS{-11} \sbcgs).

Table~\ref{t:resolved} shows the total resolved fraction based on
Tables~\ref{t:ap} \& \ref{t:count}.  For instance, with the \VF mode
cleaning using the
stowed data of Period E, the total net photons ($N\Ss{T}$) in the 6.5--7.1 keV band in the HRES region
are estimated to be 523 $\pm$ 119 (Table~\ref{t:count}). 
The resolved 356 sources by \wvdecomp (4.5$\sigma$)
contain 316 $\pm$ 24 net photons ($N\Ss{N}$) (Table~\ref{t:ap}). 
This would mean the total resolved
fraction of 61 $\pm$ 15\% (vs.~65 $\pm$ 17\% in
Table~\ref{t:resolved}), but aperture correction for missing photon
counts (loss fraction $X$) due to the finite aperture size needs to be
taken into account.  

For the fixed 2\arcsec\ radius PSF, R09 assumed $X$ to be 10\%. However,
there are caveats in this assumption when calculating the total resolved
fraction and our simulation shows the proper value is about 7\%
(Appendix~\ref{s:apcor}).  After the aperture correction, we get the total
resolved fraction of 65 $\pm$ 17\% from the 356 sources by \wvdecomp
(4.5$\sigma$) under the \VF mode cleaning, using
the stowed data of Period E (Table~\ref{t:resolved}).  For the 473
sources by R09, we get 73 $\pm$ 19\%. With the combined stowed data
set of Period D and E (see Appendix~\ref{s:inst}, we get 83 $\pm$ 22\%, which is
essentially identical to 84 $\pm$ 12\% reported by R09 (before accounting
for the unresolved CXB, which is about 4\% according to R09, see
\S\ref{s:discussion}).  However, there is a,
perhaps critical, difference in the procedure, which is the estimation
of ratio of source to background aperture regions ($r$). 
Table~\ref{t:resolved} lists $r$ for each case  (8.7\% vs.~2\%
in R09).  We will address this issue in Appendix~\ref{s:ratio} and here we
review the results in Table~\ref{t:resolved}.

Various event repositioning methods or aperture choices made little differences.
The only major difference comes from the \VF mode cleaning, which
increases the resolved fraction by about 10--12\%. The increase mainly
comes from the significant reduction in the total net photons of HRES.
For the 473 sources by R09, the \VF mode cleaning
results in 355 net photons in the source regions, which is similar to
366 net photons by the \F mode (Table~\ref{t:ap}). On the other hand, in the HRES
region as a whole, the \VF mode cleaning produces 523 net photons as
opposed to 643 photons by the \F mode.  The two results are consistent
within 1$\sigma$ due to the large errors, 
and these large statistical errors dominate the uncertainty
of the total resolved fraction.
Projecting from the count decrease by 11 in
the source region by the \VF mode cleaning\footnote{Compare the net
photons (355 vs.~366) in 6.5--7.1 keV
between 1) and 3) of the R09 column in Table~\ref{t:ap}}, 
we estimate that the total real photons removed by
the cleaning in HRES is about 19.\footnote{Using the ratio of the
total net photons in HRES (\F\ mode in Table~\ref{t:count}) to the
net photons in the source region for 3) of R09 in Table~\ref{t:ap}:
11 $\times$ 643/366.} In other words, at least about 80\% of 100 events
removed by the \VF mode cleaning are indeed background events. 

Note the total resolved fraction has a relatively larger error compared to
net photon counts in the source regions due to the additional uncertainty
of the instrumental background subtraction. The uncertainty of the
instrumental background is dominated by the relatively poor statistics
of the stowed data in comparison to the 1 Ms observation (the count
ratio in the
9--12 keV band between the two is 2.6, as shown in Table~\ref{t:count}).
The dominance of the stowed data in the error budget becomes clear
when the Period D data set is used; more than 100\% is resolved (not shown),
which is improbable (see Figure~\ref{f:radial} and Appendix~\ref{s:inst} 
for more about large variations between the two stowed data sets).

\begin{figure} \begin{center}
\includegraphics*[width=0.47\textwidth]{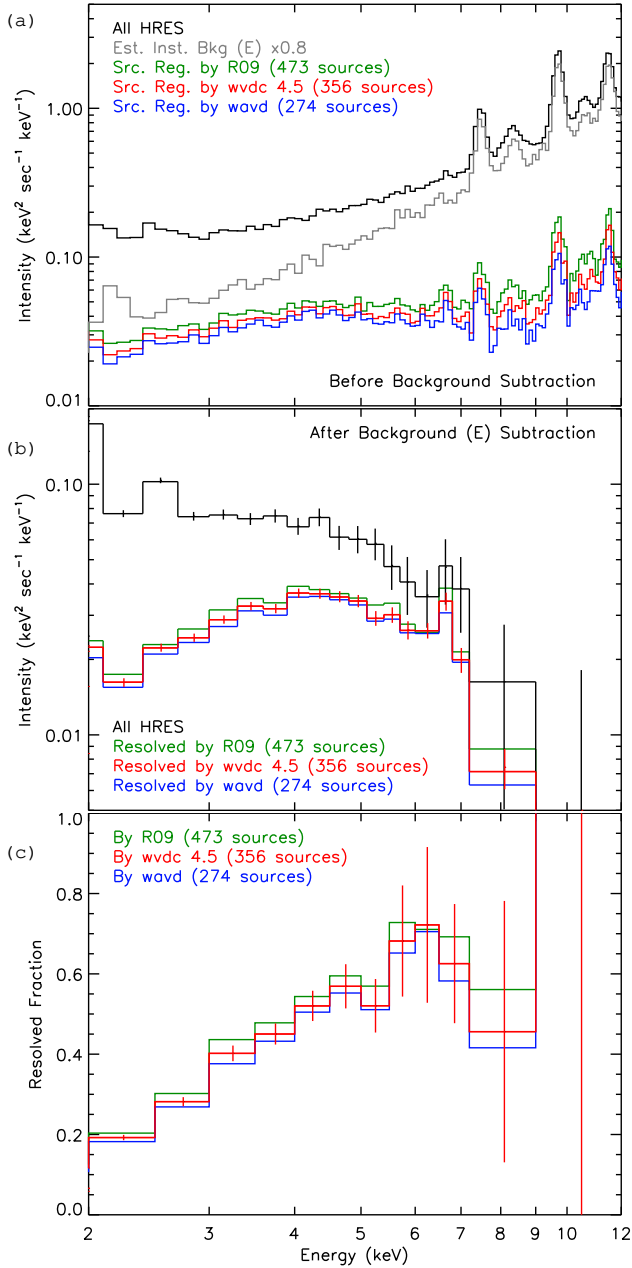}
\end{center}
\caption{GRXE spectrum and resolved fraction. (a) The X-ray spectrum before background
subtraction, (b) the X-ray spectrum after background spectrum, and (c) the resolved fraction
as a function of energies. The total spectrum in HRES and the results
for three source lists are shown. In (a), the estimated instrumental background
is shown in grey (scaled by 0.8 for clarity) and the spectra for the
source regions are not corrected for event loss due to the finite PSF size.
In (b) \& (c), the event loss is corrected (see the text). The results are
based on the default parameter choice (the EDSER routine, the \VF mode cleaning,
the Period E stowed data set, and fixed 2\arcsec radius source apertures).}
\label{f:resolved}
\end{figure}

Figure~\ref{f:resolved} shows the total net spectrum of HRES, the
resolved spectrum, and the resolved fraction as a function of energy and
compare their results with three source search routines.  
Both Table~\ref{t:resolved} and Figure~\ref{f:resolved} show that the
additional $\sim$ 100 sources added by R09 (or by the \wvdecomp routine
at 4.0$\sigma$) relative to the 356 sources from the \wvdecomp routine
at 4.5$\sigma$ do not contribute significantly to the resolved X-ray
flux in the iron emission lines and the 2--9 keV band in general.

\begin{figure} \begin{center}
\includegraphics*[width=0.47\textwidth]{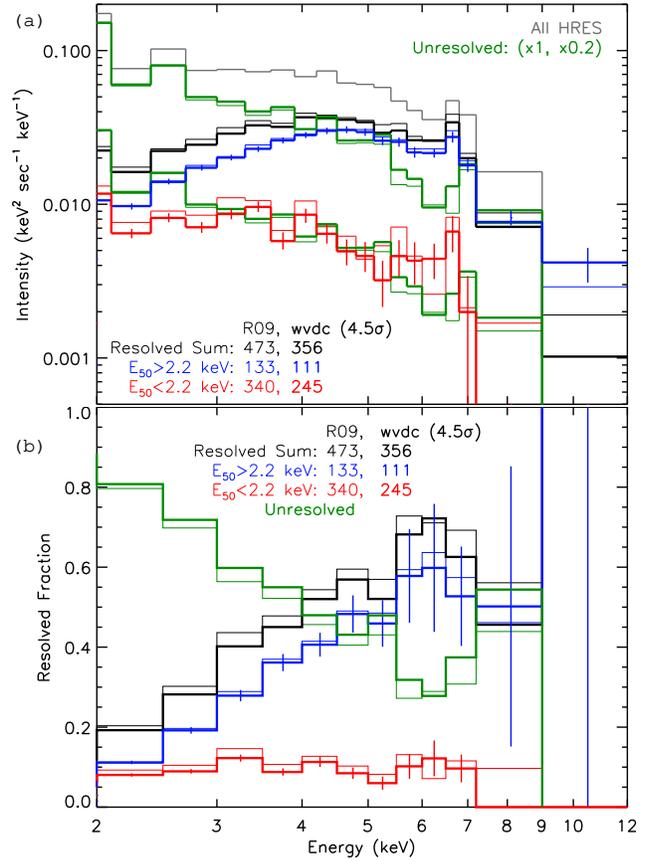}
\end{center}
\caption{The same as Figure~\ref{f:resolved} for two source lists by
\wvdecomp (4.5$\sigma$, thick lines) and R09 (thin lines). The
contribution from
the resolved discrete sources are broken up by the soft (red, $E\Ss{50}$ $<$
2.2 keV) and hard (blue, $E\Ss{50}$ $>$ 2.2 keV) sources. The residual
unresolved
X-ray spectrum and its twin scaled by 0.2 
are shown in green for comparison with the X-ray spectra from the hard
and soft sources. Among the additional 117 sources in R09 relative to
the 356 sources from \wvdecomp (4.5$\sigma$), the increase in the
resolved fraction above 5 keV is mainly from the 22 hard sources
rather than the 95 soft sources.}
\label{f:breakup}
\end{figure}

Figure~\ref{f:breakup} shows the resolved spectrum and fraction for soft
($E\Ss{50} <$ 2.2 keV) and hard ($>$ 2.2 keV) sources. 
There is a clear disparity in the combined spectrum between the
hard (blue) and soft (red) sources, which is consistent with the large variations in
count-to-flux ratio conversion factors between the different spectral
groups in Table~\ref{t:conv}.  The large majority of the resolved fraction
above 3 keV comes from the hard sources, which are likely MCVs and AGN,
whereas the soft sources such as ABs and non-magnetic CVs contribute
about 15\% or less. See also Appendix~\ref{s:hardness}.  Among the additional 117
sources in R09 relative to the 356 sources from \wvdecomp (4.5$\sigma$),
the increase in the resolved fraction above 5 keV is mainly from the 22
hard sources rather than the 95 soft sources, although the increase is within
the statistical uncertainty.
This implies that if indeed some real sources are added by
lowering the detection threshold between these two source lists, they are
mainly in the hard X-ray sources, and the trend of the dominance of
the hard X-ray sources in the 6.5--7.1 keV band continues at low fluxes.

\begin{figure} \begin{center}
\includegraphics*[width=0.47\textwidth]{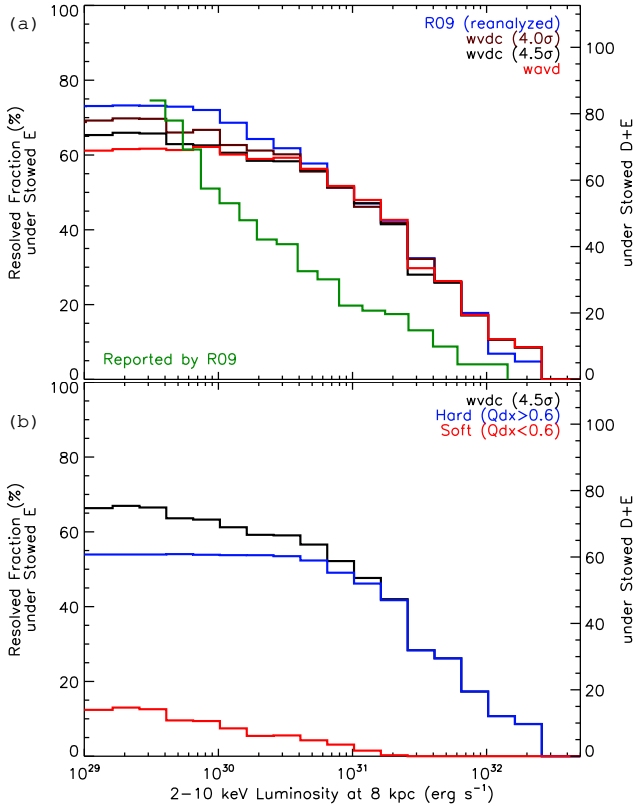}
\end{center}
\caption{The resolved 6.5--7.1 keV flux as a function of the 2--10 keV
luminosity at 8 kpc. (a) The total resolved fractions by three source
lists are compared with the result by R09. (b) The contribution from
the hard and soft sources are separated out for the 356 sources by
\wvdecomp (4.5$\sigma$).  Note the hard sources reaching the plateau at
higher luminosity is due to the higher sensitivity limit for the hard
X-ray sources (see Figure~\ref{f:qtil}).}
\label{f:resfrac}
\end{figure}

Figure~\ref{f:resfrac} shows the X-ray luminosity distribution of the
resolved fraction in comparison with the results of R09.  Our analysis
show the faint sources does not contribute significantly to the
resolved fraction (See also Appendix~\ref{s:brightness}).

\section{Discussion}\label{s:discussion}

Our estimate of the total resolved fraction ($73\pm 19$\% under Period
E or $83 \pm 19$\% under Period D+E) in the 6.5--7.1 keV band in HRES
from the 473 sources found by R09 is consistent with the estimate by R09
($84\pm 12$\% under Period D+E) within the large uncertainties. However,
our results regarding source search, aperture photometry, and the
resolved fraction, all consistently indicate that the relatively bright,
hard X-ray sources such as MCVs and AGN contribute $\sim$80\% of the
resolved flux, more dominantly than the faint, soft X-ray sources
such as ABs and non-magnetic CVs, which contribute $\lesssim$20\%.
Subsequently our results indicate the faintest $\sim$ 100 sources found
by R09 are insignificant.  Therefore, we consider the results from the
356 sources by \wvdecomp (4.5$\sigma$) is more reliable, and 
we can confidently claim that the resolved
fraction in HRES 
is $65\pm 17$\% (Period E)
or $73\pm 19$\% (Period D$+$E). Assuming the unresolved CXB in the
6.5--7.1 keV flux to be $2.9\times 10^{-13}$ \sbcgs (R09), which is about
3--4 \% of the total intensity (Table~\ref{t:count}),
the total resolved fraction
of the 6.5--7.1 keV flux in HRES is $69\pm 17$\% (Period E) or $77\pm
19$\% (Period D$+$E).  This result is also
roughly consistent with $88\pm 12$\% by R09, but unlike R09, our finding
of the dominance of relatively bright, hard X-ray sources in the resolved
fraction draw a drastically different picture in the source composition
and strongly motivate the analysis beyond HRES and the LW.


First, our finding is consistent with the previous studies indicating MCVs
are likely major candidates for the low luminosity Bulge hard X-ray sources
(e.g.~M03, M09, H09b).  These studies also found a hint of an excess
of hard X-ray sources in the central Bulge relative to stellar
population models.  
On the other hand, \citet{Revnivtsev06,Revnivtsev07,Revnivtsev09} argue
the similarity of the Galactic distribution of the GRXE and the IR flux,
which follows the stellar population.  This apparent inconsistency is
in part due to the relative shallow survey of the
previous studies ($<100$ ks).\footnote{We do not consider the possibility
of inaccurate modeling of stellar population or the disparity between
the stellar population and the IR flux, both of which are outside of
this analysis.} The excess of the hard X-ray sources is only observed
in the very central Bulge within a few arcmin of Sgr A*, given the
shallow exposure of the large survey (e.g.~M09, H09b).  The overall
X-ray flux used to match with the IR flux is still dominated by the soft
X-ray contribution, which is largely unresolved and may not trace the
hard X-ray sources.  Therefore, the hard X-ray emission of the GRXE can be
mainly from a relatively small number of hard X-ray sources, and the
above apparent inconsistency can be explained by the fact that these
previous studies were tracing a different population of X-ray sources.

Second, if indeed relatively bright, hard X-ray sources dominate the
iron emission line flux of the GRXE, we should be able to resolve a large
fraction of the same emission line in the region beyond HRES despite some
sensitivity loss due to the large offset.  This argument also applies to
the Galactic plane fields with relatively high extinction if a similarly
ultra-deep exposure is available.  Then with the majority of the hard X-ray
emission in the GRXE being resolved, we are positioned to investigate
the possible variation of X-ray source composition between the fields and
Bulge, which can provide a hint in the unique formation and evolutionary
history of the Bulge.  Extending the analysis beyond HRES also allows
a significant reduction of the uncertainty in the resolved fraction by
improving the instrumental background statistics of the stowed data set,
which is the dominant source of the uncertainties
(Appendix~\ref{s:inst})

In the following, we extend our analysis beyond HRES in light of our new
results and discuss future observations and analysis plan to improve
our understanding of the spatial variance of the source composition
and the GRXE \citep[see also][]{Morihana12}.
Finally, we summarize our thoughts on the origin of the
discrepancy between our results and R09 and some of the analysis
caveats in the Appendix (\ref{s:disc} \& \ref{s:caveats}).

\begin{figure} \begin{center}
\includegraphics*[width=0.47\textwidth]{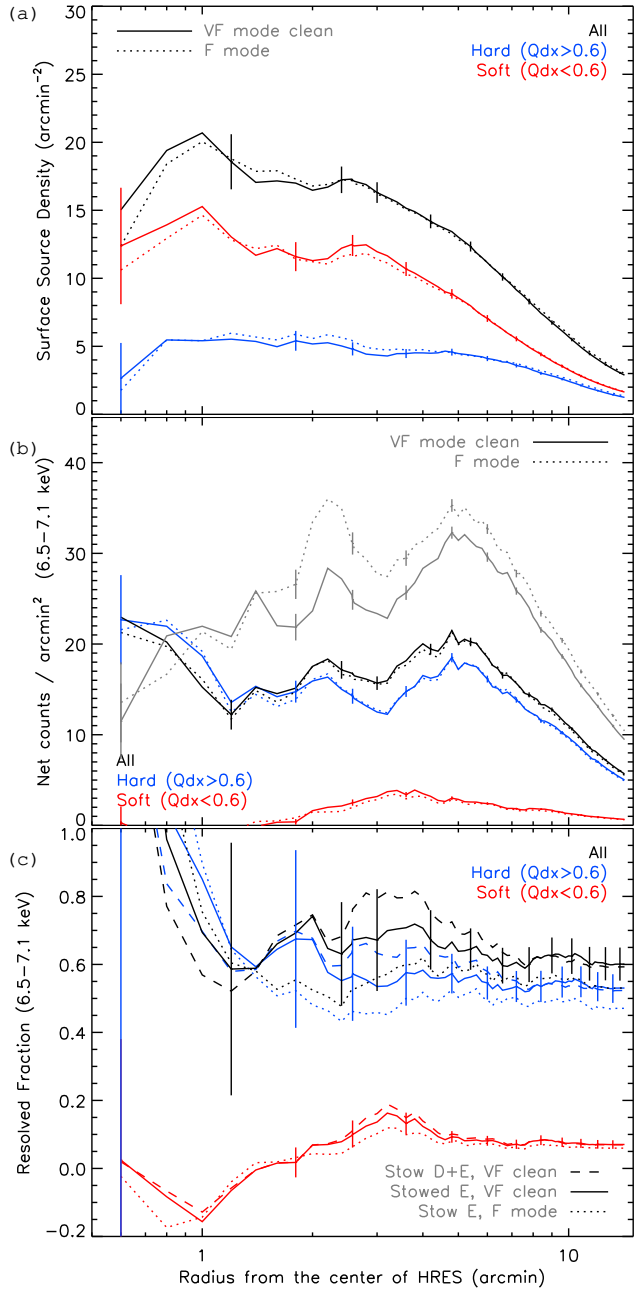}
\end{center}
\caption{Radial distribution of (a) the source surface density, (b)
the net photons in the 6.5--7.1 keV band within the given radius, and (c)
the resolved fraction using the sources detected by \wvdecomp (4.5$\sigma$). Source apertures are 1.5 keV 95\% PSF and the
background apertures are the surrounding annulus (2\x and 5\x of the
PSF radius for inner and outer radii respectively). 
The solid lines are for \VF mode cleaning and the dotted lines for \F mode. 
The soft (red) and hard (blue) sources are separated based on the
median energies.} \label{f:radial}
\end{figure}

\subsection{Beyond HRES} \label{s:impl}

\begin{figure} \begin{center}
\includegraphics*[width=0.47\textwidth]{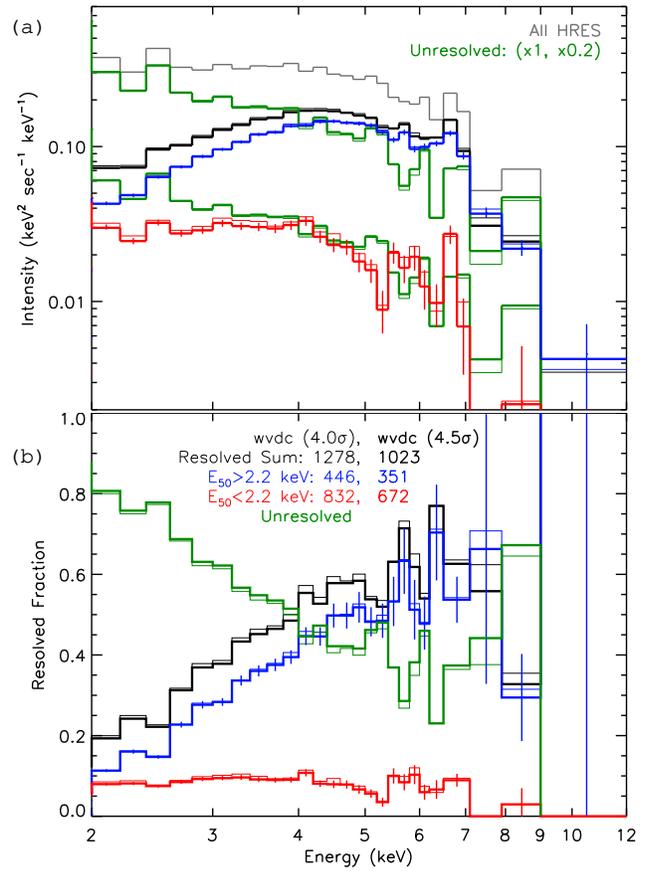}
\end{center}
\caption{The GRXE and resolved spectra (a), and the resolved fraction (b) 
in the central region within the 5\arcmin\ radius (using the Period D$+$E stowed data set). The results are
shown for two source lists by \wvdecomp (4.0$\sigma$ for thick lines
and 4.5$\sigma$ for thin lines). The grey line is for the GRXE spectrum,
and the black lines are the resolved X-ray spectrum. The blue and red lines are
the resolved spectra from the hard and soft sources respectively. 
The green lines show the unresolved spectra and their scaled twins (\x0.2)
for comparison with the combined spectra of the soft X-ray sources.
Compare this with Figure~\ref{f:breakup}.}
\label{f:breakup2}
\end{figure}

Figure \ref{f:radial} shows the source number density, net photon counts,
and the resolved fraction of the 6.5--7.1 keV flux as a function of
the radius from the center of HRES using the resolved sources from the \wvdecomp routine (4.5$\sigma$). The figure also shows the hard
(blue) and soft (red) sources separately, and the results with (solid)
and without (dotted) the \VF mode cleaning. For a given source, we use the
1.5 keV 95\% PSF for the source region and the surrounding annulus for
the background region for aperture photometry since the fixed
2\arcsec\ radius aperture is no longer applicable at large offset angles and using the rest of HRES
(excluding the source regions) for the background region does not
reflect the local variation of background. The use of variable
apertures is justified since in HRES both cases of the aperture
choices (variable and fixed sizes) produce the consistent results
(Table~\ref{t:ap} Options 1 vs.~2).

As expected, the surface density of the soft sources (red in
Figure~\ref{f:radial}a) show a more
significant change with the radius
of the analysis region than that of the hard sources (blue in
Figure~\ref{f:radial}a). The former drops noticeably
beyond 2.5\arcmin\ from the center, whereas the latter are
more or less uniform out to 5\arcmin.
The similar trend can be seen in the total
net photons in the 6.5--7.1 keV band from the hard X-ray sources (blue in
Figure~\ref{f:radial}b) and the analysis region altogether (grey in
Figure~\ref{f:radial}b), which peaks at around 5\arcmin.  On the other
hand, the 6.5--7.1 keV photons from the soft X-ray sources (red in
Figure~\ref{f:radial}b) rises up and peaks at 3.5\arcmin\ and drops
afterwards, which can be partially explained by a patch of seemingly
diffuse soft emission region at around 3--4\arcmin.

The contribution to the resolved fraction from the soft sources are
insignificant throughout the field. In fact their contribution in the
very central region is non-existent despite the relatively large number
of sources found in the region. This again supports our finding of the
dominance of the relatively bright, hard X-ray sources in the resolved
GRXE. It is also consistent with a recent independent analysis of the
region by \citet{Morihana12}.
In the case of the resolved fraction, the radial variation is not
significant, but its uncertainty drops noticeably as extending to the
larger region.  If we limit our analysis to 5\arcmin\ where the total
net photon density in the 6.5--7.1 keV band of the analysis region is
highest, the resolved fraction is 64 $\pm$ 6\% (Period E) or 69 $\pm$ 7\%
(Period D$+$E) for \wvdecomp (4.5$\sigma$); after including the 3.7\% CXB
contribution, the total resolved fraction is 68 $\pm$ 6\% (Period E) or 73
$\pm$ 7\% (Period D$+$E).\footnote{The total resolved fraction does not change
significantly even out to 10\arcmin. This is because the X-ray flux at
large off-axis angles does not contribute significantly due to the
reduction in the effective area. Subsequently there is no improvement in the
uncertainty of the total resolved fraction.}

\subsection{Unresolved GRXE} \label{s:unresolved}

Figure~\ref{f:breakup2} shows the GRXE and its resolved
spectra in the 5\arcmin\ radius region. 
The result is consistent with that in HRES (Figure \ref{f:breakup}).
Now with higher statistics, one can see a few more features in the
spectra.  First, there appears to be a lack of the 6.4 keV emission line
in the region (e.g.~see the black lines in the 6-6.5 keV band in
Figure~\ref{f:breakup2}a). Although 
the ACIS CCD spectral resolution in HRES may
not be suitable for clear separation of the 6.4 and 6.7 keV lines
under the given relatively poor statistics \cite[cf.][]{Ebisawa08},
other Bulge fields such as the Galactic center strip surveyed by
\citet{Wang02} show a prominent emission line feature at 6.4 keV,
a large fraction of which may be of diffuse origin.
Second, the spectra from the resolved sources, in particular,
the soft sources, shows an absorption feature in the 5--5.5 keV band.
In turn, the unresolved spectrum (green) shows an emission feature in
the same energy band.  

Given high stellar density in the LW, where often more than a few stars
with $V\lesssim24$ within the error circle of X-ray positions are observed
in the \HST image (V09), it is still possible that the unresolved 20--30\%
can be from the X-ray emission of the unresolved discrete sources.
The relatively soft X-ray spectrum of the unresolved residual GRXE (green
in Figures~\ref{f:breakup} \& \ref{f:breakup2}) suggests a possibility that the unresolved
discrete sources are mainly soft coronally active stars such as ABs.
In fact, the unresolved spectrum is clearly softer than the
combined spectrum of the soft sources, so there may be more contribution
from ABs than non-magnetic CVs in the unresolved spectrum, whereas
the resolved spectrum of the soft sources may have relatively larger
contribution from non-magnetic CVs.  This interpretation is consistent
with ABs and other coronally active sources being fainter ($\sim
10^{28-31}$ \lcgs) than accreting sources such as non-magnetic CVs
($\sim 10^{29-32}$ \lcgs), but it also implies the contribution from
ABs is not resolved at this luminosity limit.

Alternatively it is also possible that the hard X-ray flux ($>$ 4 keV)
of the unresolved GRXE is mainly from the faint hard X-ray sources such
as MCVs and the soft flux from the coronally active stars like ABs.
For instance, the X-ray spectrum of the unresolved CXB ($\sim$ 4\% of
the total flux in the 6.5--7.1 keV band) can be described by a power
law spectrum with photon index of 1.4.  In other words, the combined unresolved spectrum
(green) in Figures~\ref{f:breakup} \&
\ref{f:breakup2} contains a contribution from the sources whose
spectra are much harder than the combined unresolved spectrum itself
or the combined spectrum of the soft sources.
If the 8\% increase in the resolved
fraction of the 6.5--7.1 keV band from the 356 sources by \wvdecomp
($4.5\sigma$) to the 473 sources by R09 is credible (thick
to thin lines in Figure~\ref{f:breakup}), the latter scenario is supported
by the fact that most of the 8\% increase is from the faint hard X-ray
sources but not from the faint soft X-ray sources.  Then the paucity of
the hard X-ray sources implies a truly diffuse hard X-ray component
may be present
in the GRXE. 

\subsection{Future Studies} \label{s:future}

It is now possible to draw a rather complete picture of
the Galactic X-ray source composition and their Galactic distribution,
through resolving the majority of the GRXE by
ultra-deep \chandra exposures. This calls for more observations and analysis
of the other fields.  The LW field, while perhaps ideal for resolving the GRXE
due to the proximity to the GC and the low extinction, may not represent a
typical Bulge field in the Galactic Plane.  
First, the
total X-ray spectrum of the LW field lacks the neutral Fe 6.4 keV emission
line, which is often outstanding in the Plane fields and suspected
to be mainly from the diffuse emission \citep[e.g.][]{Wang02}.  Therefore,
an ultra-deep exposure of the Plane fields (apart from the Sgr A* field,
which contains many complex diffuse features) is required.  Second,
unlike the Fe 6.7 keV line, the 2--6 keV medium-hard flux remains
largely unresolved.  As seen in the previous section, comparison of
the unresolved GRXE spectrum with the combined spectra of the soft and
hard X-ray sources may indicate spectral transitions from hard MCVs,
to soft non-magnetic CVs, and to even softer, unresolved ABs (or diffuse
components).  Therefore, the medium band GRXE and its spatial variation
will allow modeling of the relative composition of the three major source
types.  For this, another low extinction field such as Baade's Window
(BW) at 4\Deg south of the GC might be ideal.  For instance, unlike
the LW, in BW the apparent diffuse X-ray emission is remarkably absent.
The lower extinction and the lack of the apparent diffuse emission in BW
improve a chance of resolving the GRXE in a broader band than in the LW.
In addition, unlike the hard X-ray sources whose density falls radially
from the GC, there is an excess of the soft X-ray sources in BW relative
to the LW at the same 100 ks exposure (H09b), despite the larger offset
of BW from the GC (4\Deg vs.~1.4\Deg for the LW).  Therefore, when the
spatial variance of the GRXE between BW and the LW is compared to the
variance of the hard and soft X-ray source numbers, the unambiguous
contribution of each source type to the GRXE can be calculated.

The wide band coverage (5--200 keV) and large effective area ($\gtrsim$
700 cm\sS{2} at 7--12 keV) of the Nuclear Spectroscopic
Telescope Array (NuSTAR) \citep{Harrison05} bring a new promise of constraining the GRXE.
A mildly deep observation (e.g.~200 ks\footnote{This is actually much
deeper in the hard X-ray band ($>$7 keV) than the 1 Ms \chandra exposure
since the effective area  of \chandra is less than 100 cm\sS{2}
above 7 keV.}) of the LW with NuSTAR will enable the absolute intensity
measurement of the GRXE in the region above 6 keV. Such a measurement will
allow a precise calculation of the resolved fraction of the GRXE by the
\chandra sources without relying on the somewhat uncertain \chandra/ACIS
instrumental background.

\section{Summary} \label{s:summary}

Through an independent analysis of the X-ray sources in the LW, we
resolved the iron emission line of the GRXE in the 6.5--7.1 keV band up
to (69--77) $\pm$ 19\% in the central circular region of 2.56\arcmin\
radius\footnote{This is from \wvdecomp (4.5$\sigma$) under the \VF clean mode in
Table~\ref{t:resolved}, including $\sim$ 4\% of the unresolved CXB.
For singling out this result, see \S\ref{s:discussion}.}
and (65--73) $\pm$ 7\% for the 5\arcmin\ radius.
The dominating uncertainty is from
the instrumental background in both statistical and systematic nature
(e.g.~\VF mode cleaning, Period D$+$E vs.~E).  We find that the resolved
GRXE is
dominated by the relatively bright ($\gtrsim$$10^{31}$ \lcgs), hard
X-ray sources ($E_{50}$$\gtrsim$2.2 keV), which are likely MCVs and
AGN. The relatively faint, soft X-ray sources such as ABs and non-magnetic
CVs do not contribute more than 20\% of the resolved flux.  The refined
resolved fraction in the 5\arcmin\ radius region leaves room for truly
diffuse components in the GRXE, but the undetected
large population of the relatively faint ($\lesssim$$10^{31}$ \lcgs), hard
X-ray sources can make up for the unresolved
fraction. We also believe we have identified a few analysis caveats
in R09, which led to the disagreement with our
results regarding the source composition of the resolved GRXE.

\section{Acknowledgement} 

We thank M.~van den Berg and M.~Servillat for reading the manuscript
and useful comments. We thank V.~Kashyap for his help on calculation
of detection confidence. We also thank M.~Revnivtsev for providing his
source list and the extensive discussion on the topic and analysis despite
some disagreement in the results.  We also thank J.~Grindlay
for his support and useful suggestions in the analysis.

\appendix

\section[]{Detection Significance} \label{s:detsig}

\begin{figure} \begin{center}
\includegraphics*[width=0.48\textwidth, clip=true,trim=0mm  0mm -0mm 0mm]
{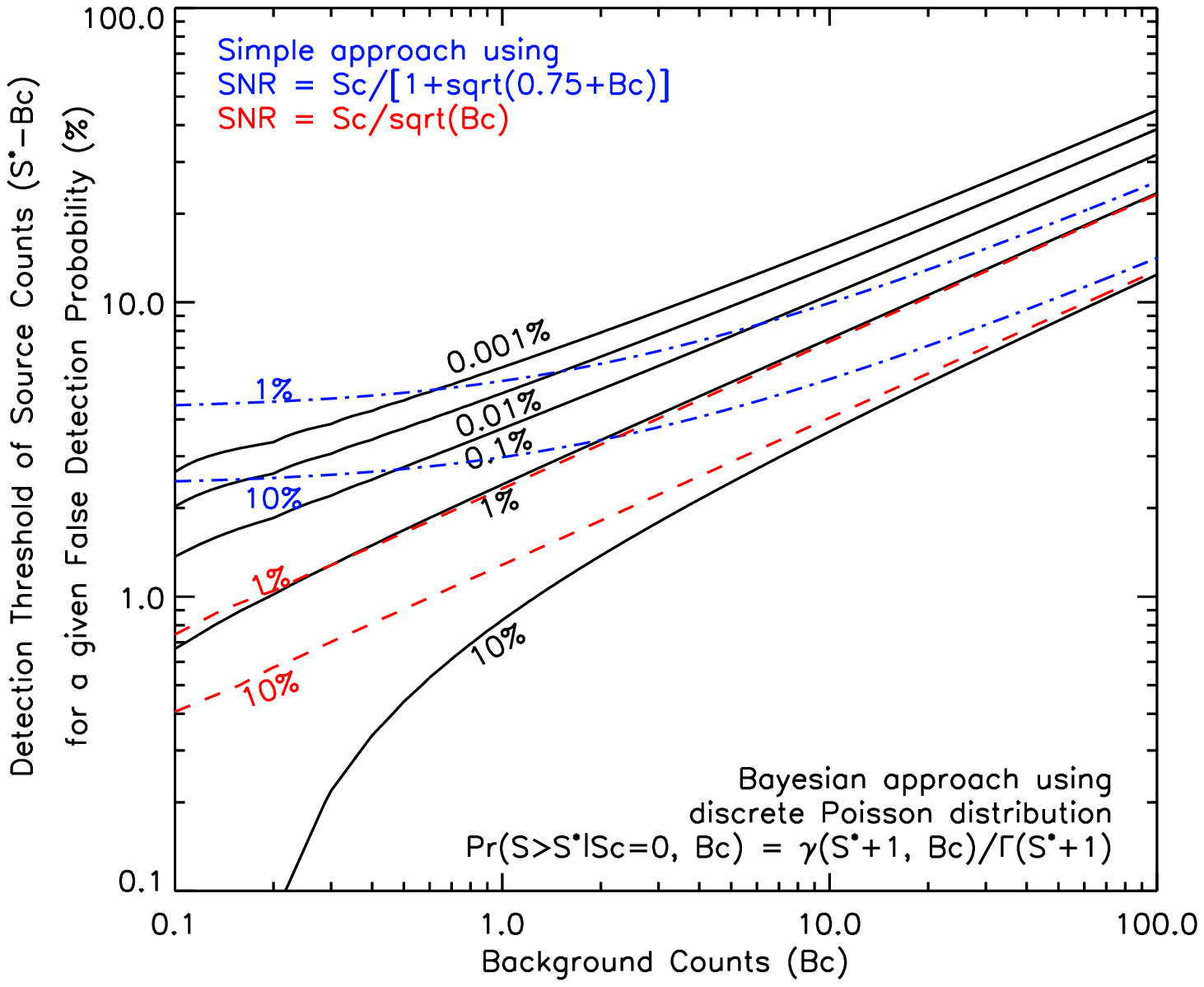}
\caption{Detection threshold ($S^*-B_C$) for source count under a given
false detection probability ($P_r = 1-C$, where $C$ is the
detection confidence).  The simple minded approaches using 
signal-to-noise ratios (SNRs)
(red dash for Gaussian statistics \& blue dotted-dash for Gehrels'
approximation for Poisson statistics) is compared with the Bayesian 
approach (black solid) by
\citet{Weisskopf07, Kashyap10}. Note $P_r$ from the Bayesian analysis
is defined for the total counts ($S$) in the detection cell (see Figure~2 in
\citet{Kashyap10}, so technically the Bayesian approach does not provide
the detection threshold for source counts ($S_C$), but for easy
illustration of how bright the source needs to be for detection, the
detection threshold is expressed for $S_C$ under the assumption that $B_C$
is known (subsequently the plot ignores the discrete nature of the
counts as well). Ironically using the SNR based $P_r$ under Gehrels'
approximation without accouting for trial statistics can underestimate
the false detection probability for high background cases.}
\label{f:detsig}
\end{center}
\end{figure}

Detection significance is often estimated based on signal-to-noise ratio
(SNR). If the background count ($B_C$) in the detection cell ($\Omega_T$)
is known, the SNR to describe detection significance\footnote{Note the
difference from the SNR often used to describe the
confidence range of the source count, where the noise term includes
the error contribution of
the source count as well. e.g.~when the error of the background count
is estimated independently, SNR =
$S_C$/($\sigma_{T_C}^2+\sigma_{B_C}^2)^{0.5}$ 
where $T_C$ is the error of the total counts in the source region.} is
given as $S_C$/$\sigma_{B_C}$, where $S_C$ are the photon counts from the source
and $\sigma_{B_C}$ is the fluctuation of the background counts.
For Poisson distribution, $\sigma_{B_C}$ is simply $B_C^{0.5}$ or
$1+(0.75+B_C)^{0.5}$ at low counts ($\lesssim$ 15) \citep{Gehrels86}.
The SNR-based approaches involve a number of approximations: for
a given confidence ($C$) or a false detection probability ($P_r =
1-C$), it is often assumed that the SNR follows a Gaussian distribution.
For instance, for $>$90\% confidence ($P_r < 0.1$) with $B_C = 1$, $S_C >
1.28$ (SNR $>1.28$) or $S_C > 2.97$ for Gehrels' approximation, see
red and blue lines in Figure~\ref{f:detsig}).

A more rigorous approach based
on discrete Poisson distributions and a Bayesian treatment of false
detection probability can be found in the literature: 
Eq.~A11 in \citet{Weisskopf07} (see also Eq.~A8) and footnote 13 in
\citet{Kashyap10}.  Their formulae are identical\footnote{Note that
there is an error in the formula in footnote 13 of \citet{Kashyap10}:
$\gamma(S^*+1, \tau_S \lambda_B) = \int_{0}^{\tau_S \lambda_B} e^{-t} t^{n_S} dt$.}
to each other except for
their interest of unmarginalized parameters: the cell size ($\Omega_T$)
in the former and the exposure ($\tau_S$) in the latter.  If we assume
$B_C$ is measured, the false detection probability is simplified as
\begin{eqnarray*}
	P_r(S>S^*|S_C=0,B_C) & =&  1 - C	\\
		& =&  \displaystyle\sum\limits_{m=0}^{S*} \frac{{B_C}^m}{m!}e^{-B_C} \\
		& =&  \frac{\gamma(S^*+1,Bc)}{\Gamma(S^*+1)},
\end{eqnarray*} where $\gamma$ and $\Gamma$ are incomplete and regular
gamma functions respectively.
Under this approach, for $>$ 90\% confidence with $B_C=1$,
$S_C>0.8$\footnote{Note this is not entirely a correct statement since
$P_r$ is defined for $S$, the total counts in the source region, but
not for $S_C$, the source count.  The correct statement is $S > 1.8$.}
(see black lines in Figure~\ref{f:detsig}). 

We use a circular region of 1\arcsec\ radius around each source for
detection cell, following \citet{Weisskopf07}, and calculate the source
counts ($S_C$) by subtracting the background counts ($B_C$) from the
total counts ($S$) in the cell.  For the background counts ($B_C$),
we take the counts in an annulus around the source with 4\arcsec\ and
10\arcsec\ radii (excluding the 3\arcsec\ radius circles of neighboring
sources), and scale them by the ratio of the detection cell size to the
background region.  Here we assume that $B_C$ represents the true mean
value of the background counts in the detection cell for simplicity.
The range of $B_C$ for the sources in HRES is 3.1 to 21.  Note the
detection cell is chosen to be smaller than the source aperture regions
in aperture photometry (\S\ref{s:ap}), since the former is designed for
efficient source detection and the latter is designed for accurate 
flux estimation.

Figure~\ref{f:detsig} compares the two approaches under the assumption
that $B_C$ is known.  
Gehrels' approximation 
is often used to account for the asymmetric deviation of Poisson
distributions from Gaussian distributions at low counts ($\lesssim$ 15). 
However, Figure~\ref{f:detsig} shows in fact using simple Gaussian errors
is more accurate than using Gehrels' approximation, indicating the latter
may result in detection loss of faint sources.  On the other hand. the
real data can often deviate from a pure Poisson distribution or contain
features that are not easy to account for, and thus it is usually a
safe approach to have a higher threshold by using Gehrels' approximation.

For detection confidence of a population of sources discovered by a
search routine, one has to take into account the number of search trials
explicitly.  Without accounting for trials statistics, both simple
Gaussian errors and Gehrels' approximation of Poisson errors result
in a wrong estimate of false detections.  Ironically using Gehrels'
approximation without accounting for trial statistics may accidentally
produce a proper estimate of false detections at low background count
cases but it will underestimate false detections at high background count cases.
e.g.~in Figure~\ref{f:detsig}, the detection threshold for $P_r$=1\%
using Gehrels' approximation (the blue line) matches the threshold
for $P_r$=0.01\% from the Bayesian approach (the black line) when
$B_C\sim1$, but the threshold for the former is lower than the latter
when $B_C\gtrsim2$.

\section[]{Origin of Discrepancy in Aperture Photometry} \label{s:disc}

In our opinion inaccurate estimations of the following three quantities
in R09 are the major origins of the discrepancy between R09 and ours in
the aperture photometry results.

\subsection{Source to Background Region Ratio} \label{s:ratio}

Table~\ref{t:resolved} shows the exposure map corrected geometric sky
ratio ($r$) of the source to background aperture regions after overlap
correction.  The ratios gradually increase from $\sim$ 5\% to 9\% as the
source numbers increase from the 274 to 473 sources.  For comparison, the
473 fixed 2\arcsec\ radius circles in a 2.56\arcmin\ radius circle means
$r$=8\% without considering source aperture overlap (about 100 sources)
and the gaps between the CCDs.  We validate our calculation of the ratios
by the fact that the same aperture photometry produces the essentially
null net photons in the 9--12 keV band in the combined source regions
(Table~\ref{t:ap}).

Interestingly R09 quote 2\% for this ratio for their 473 sources, which
is a factor of four smaller than our estimate.
R09 justifies their ratio based on a
claim that their aperture photometry is done using the pixellated image
rather than event files \citep{Revnivtsev12}.
However, as shown in \citet{Li04}, we believe aperture photometry
benefits substantially from the sub-pixel information by using event
files instead of pixellated images.  In addition, a few techniques have
been developed and proven to reduce pixellation-induced uncertainties
for aperture photometry using event files.
\footnote{For instance, the exposure-corrected aperture area of a source
region is calculated by multiplying the mean value of the exposure map
in the source region with the geometric aperture size instead of adding
up the exposure map values of the pixels inside the source region. The
latter is subject to pixellation-induced errors when the aperture
size is small, whereas the former is accurate even if the aperture radius
is similar to a pixel size.
See \citet{Kim04} and H05.}

Given the dominance of the background in the region,  the underestimated ratios ($r$)
by a large factor has significant
consequences.  First, it mistakenly increases the resolved flux.
Second, it smears the spectral diversity of sources by adding a
constant term of the background spectra.  Now the effect gets amplified
proportionally to the number of sources since each source adds a constant
background contribution into its spectrum.  This generates an illusion
of increase in the resolved flux as one approaches the faint side of
flux where an increasingly large number of sources are added to the
source list. Therefore, we believe the underestimation of the ratios
($r$) contributes to the discrepancy in the X-ray luminosity distribution
of the resolved GRXE. It also explains the apparent large contribution
of the soft (relatively faint) sources to the resolved fraction of the
6.5--7.1 keV band in R09.

\begin{figure} \begin{center}
\includegraphics*[width=0.48\textwidth]{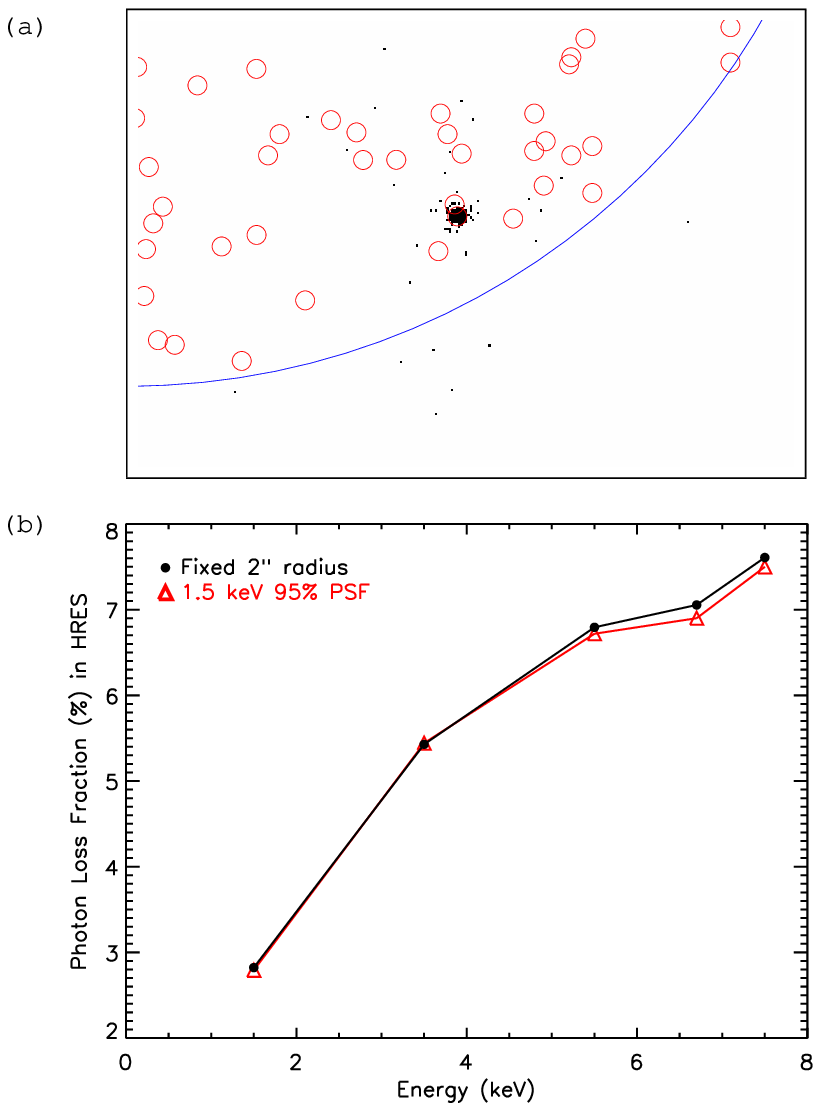}
\caption{Photon loss due to finite aperture size. (a) an example of
MARX simulations using 6.7 keV photons for a source near the boundary of the HRES region.
Some photons outside the 2\arcsec\ radius aperture of the
source still fall inside of other source regions (small red circles)
or outside of the HRES region altogether, which is outlined by a part
of the large (blue) circle.  (b) Effective photon
loss fraction of the 473 sources by R09 in HRES as a function of
X-ray energies, which is smaller than 10\% expected for a
single source in the full ACIS-I field of view (see the text).
The larger variable aperture using 1.5 keV 95\% PSF captures slightly more
photons at high energies than the fixed 2\arcsec radius apertures.
}
\label{f:loss}
\end{center}
\end{figure}

\subsection{Aperture Correction for Missing Photons} \label{s:apcor}

When the ratio of source
to background regions is $r$, the true net photons ($N_C$)
from the source is given as 
$N\Ss{C} = (N\Ss{S} - r N\Ss{B})/(1 - X - r X) = N\Ss{N}/(1- X - r X)$, 
where $N\Ss{S}$ and $N\Ss{B}$ are the total events in the source and
background regions respectively, and $X$ is the missing photon fraction
due to the finite aperture size.  For the fixed 2\arcsec\ radius PSF,
R09 assumed $X$ to be 10\%.  This is alright in estimating the true net
flux of a single point source in the field, but for calculation of the
total resolved fraction in a small section of the field, there are two
caveats as illustrated in
Figure~\ref{f:loss}. First, the multiple source regions capture more
photons than the 90\% of photons enclosed by the single source aperture.
Second, the large tail of the PSF (partially due to CCD readout intervals)
makes the source photons scatter
even outside of the HRES region. i.e. some fraction of the missing 10\%
photons are outside of the HRES region altogether,
which should not be counted for calculating the total resolved fraction
in the HRES region.  The latter is prominent for sources that fall near
the edge of the HRES region. For the proper aperture correction,
we have conducted a set of MARX simulations
using various source and background regions.
Figure~\ref{f:loss}b summarizes the effective photon loss fraction
as a function of energies for two different aperture choices using the
positions of the 473 sources detected by R09. For example, 
the proper correction factor X is 7.1\% for the fixed
2\arcsec\ radius apertures of the 473 sources in the HRES region at 6.7
keV, and 6.9\% for 1.5 keV 95\% PSF variable apertures.  This correction
factor approaches to the expected 10\% as we expand the analysis
region beyond HRES. 
Since the subsequent correction factor for the total resolved fraction is
proportional to the resolved flux before the correction, so the error
in the missing photon fraction can also get amplified accordingly.

\subsection{Count Rate to Flux Conversion Factor} \label{s:conv}

R09 use $(a + b S_C^{0.8}) S_C$ for the count-to-flux conversion factor,
where $S_C$ is the net source counts in 0.5--7 keV and $a$ \& $b$ are
constant \citep{Revnivtsev12}. This conversion solely relies on the net
counts, disregarding the spectral variation of sources in the same
count range, and the resulting $S_C^{1.8}$ term artificially stretches
the luminosity range. For instance, for two bright sources with $S_C$$\sim$300
and 800 in HRES that show similar spectral type under a simple powerlaw fit,
it is reasonable to estimate their flux would also differ
by a factor of 2.7 (=800/300), but under the above conversion scheme by
R09, their flux turns out to differ by a factor of 5.  In our opinion,
the conversion scheme by R09 likely misassigns the flux values of
many sources, and in the resulting X-ray luminosity distribution the
contribution of the faint sources appears larger than what it should be.

\section[]{Analysis Caveats} \label{s:caveats}

\subsection{Statistical Uncertainty of Stowed Data in HRES} \label{s:inst}

\begin{figure} \begin{center}
\includegraphics*[width=0.45\textwidth]
{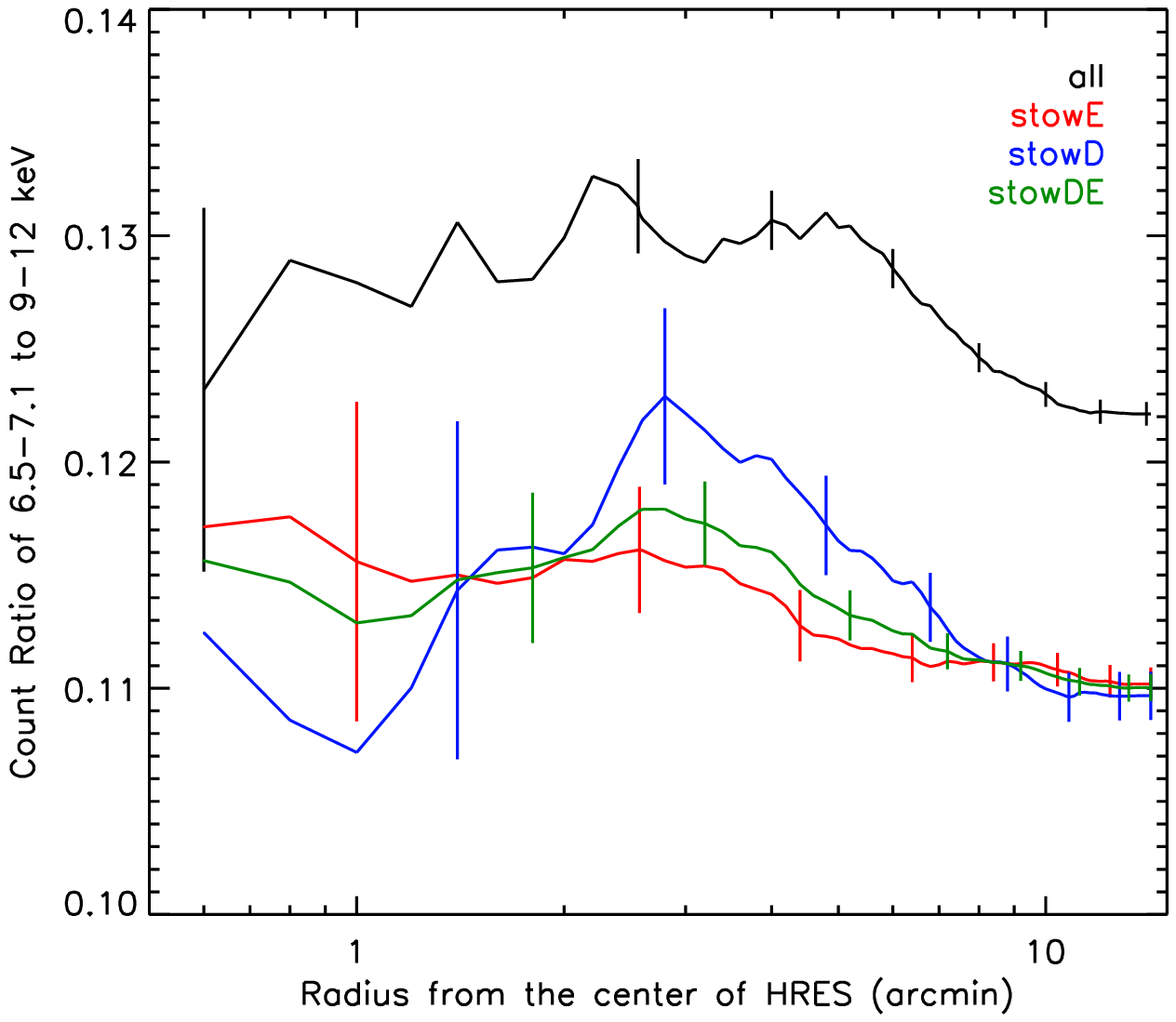}
\caption{The 6.5--7.1 to 9--12 keV count ratio as a function of the
enclosing radius from the center of the HRES. There is a significant fluctuation
at small radii. Two stowed data sets show a large difference at small radii, but
when most of the data are included, the ratios of the two data set are
consistent with each other as shown in \citet{Hicox06}.}
\label{f:fraction}
\end{center}
\end{figure}

The dominant contribution to the uncertainty of the total resolved
fraction of the 6.5--7.1 keV flux is from the statistical uncertainty
of the reprojected stowed data. In addition, there appears to be an even
larger systematic uncertainty between the Period E and D data sets.
How can this be since \citet{Hicox06} demonstrated the stowed data do not
exhibit any significant variation in the flux and spectrum over the years?
Figure~\ref{f:fraction} illustrates the origin, which plots the relative
count ratio of the 6.5--7.1 to 9--12 keV bands as a function of the
radii from the center of HRES.  When most of the data in the ACIS-I chips
from CCD 0 to 3 are used, which corresponds the right side of the plot
(radius $\gtrsim$ 8\arcmin) the relative count ratio of 6.5--7.1 to
9--12 keV do not show any significant variation between
the two periods, which is consistent with \citet{Hicox06}. But the same ratio
shows large fluctuations at small radii.

In addition, for instrumental background subtraction, the stowed data are
reprojected to sky according to the aspect solution of the observations
as aforementioned in order to properly account for the spatial variation.
Each reprojection produces different results originating from random
assignment of events to the aspect solution, and each reprojected stow
data shows a similar scale of fluctuation as the error bars of statistical origin in
Figure~\ref{f:fraction}. Note that the curves for the stow data
in Figure~\ref{f:fraction} are based on the average values of
100 separate reprojections for each data set.  Therefore, using a
particular reprojection result may lead to a significant different
outcome in the resolved fraction. 

In summary, the instrumental background is sensitive to the choice of
the stowed data set (D$+$E vs.~E), \VF mode cleaning, and the analysis
region size (unless it is larger than 7-8\arcmin\ radius). Each of these
can make the total resolved fraction vary by about 10\% or more.

\subsection{Soft vs. Hard X-ray Sources}\label{s:hardness}

One may argue the distinction of hard and soft X-ray sources by 
median energy of 2.2 keV and the subsequent association with source
types like MCVs and ABs are too crude. In fact, coronally active stars
exhibit spectral hardening during flares. However, a few dozen sources
with some clues
about their source type  (e.g.~Figure~\ref{f:qtil}) are consistent with our
division scheme, and the median energy of typical flares from coronally
active stars are below 2.2
keV (e.g.~$E\Ss{50} >2.2$ keV means $kT>$ 10 keV for thermal plasma
models with \nHt$\sim$ 0.7). Therefore, despite its arbitrary aspect,
our distinction of source types with the median energy is statistically
justified. 

\subsection{Faint vs. Bright X-ray Sources} \label{s:brightness}

One may claim the dominance of the relatively bright ($\gtrsim$$10^{31}$
\lcgs), hard ($E\Ss{50}$$\gtrsim$ 2.2 keV) X-ray sources in the resolved GRXE is because we have not
resolved the flux from the faint sources although we resolved the flux of
the bright sources properly.  If we simply add the resolved fraction of
the faint sources by R09 to the resolved fraction of our bright sources,
the total resolved fraction exceeds 100\%.  This means there should be an
error in the aperture photometry.  However, it is hard
to imagine any mistake in aperture photometry would channel the X-ray
flux into a smaller number of sources.  Errors in the analysis usually
influence the results in an opposite way, smearing the outcome rather
than sharpening the results.  For instance, the incorrect boresight
correction or lack thereof would smear the image, resulting more evenly
spreaded events among the sources and background region. In summary, the
total resolved fraction from our analysis and their dependence of the
source type and luminosity strengthen the earlier conclusion that the
majority of the resolved flux are from relatively bright, hard sources
such as MCVs and bright AGNs.


\end{document}